\documentclass[aps,preprint,floats,showpacs]{revtex4} 

\usepackage{graphicx}

\begin{document} 

\newcommand{\be}{\begin{equation}}
\newcommand{\ee}{\end{equation}}
\newcommand{\bea}{\begin{eqnarray}}
\newcommand{\eea}{\end{eqnarray}}


\title{
Amplitude-equation approach to spatiotemporal dynamics
of cardiac alternans   
}
\author{Blas Echebarria}
\affiliation{Departament de F\'{\i}sica Aplicada, Universitat 
Polit\`ecnica de
  Catalunya, Av. Dr. Mara\~n\'on 44, E-08028 Barcelona, Spain.}
\author{Alain Karma} 
\affiliation{Department of Physics and Center for 
Interdisciplinary Research on 
Complex Systems,\\
Northeastern University, Boston, MA 02115.}

\date{October 11, 2007}

\begin{abstract}
Amplitude equations are derived that describe the spatiotemporal
dynamics of cardiac alternans during periodic pacing of one- 
[B. Echebarria and A. Karma, Phys. Rev. Lett. {\bf
88}, 208101 (2002)] and two-dimensional homogeneous tissue 
and one-dimensional anatomical reentry in a ring 
of homogeneous tissue. These equations provide a
simple physical understanding of arrhythmogenic
patterns of period-doubling oscillations of action potential
duration with a spatially varying phase
and amplitude, as well as explicit quantitative
predictions that can be compared  
to ionic model simulations or experiments. The form
of the equations is expected to be valid for a large
class of ionic models but the coefficients 
are derived analytically only for a two-variable ionic model
and calculated numerically for the original Noble model of Purkinje fiber
action potential.
In paced tissue, this theory explains the formation of ``spatially discordant 
alternans'' by a linear instability mechanism that produces a periodic pattern of 
out-of-phase domains of alternans.
The wavelength of this pattern, equal to twice the spacing
between nodes separating out-of-phase domains, is shown to depend 
on three fundamental lengthscales that are 
determined by the strength of cell-to-cell coupling and  
conduction velocity (CV) restitution. 
Moreover,
the patterns of alternans can be either stationary, with
fixed nodes, or traveling, with moving nodes and hence quasiperiodic
oscillations of action potential duration, 
depending on the relative strength of
the destabilizing effect of CV-restitution and the 
stabilizing effect of diffusive coupling. For the ring
geometry, we recover the results of Courtemanche, Glass and
Keener [M. Courtemanche, L. Glass, and J. P. Keener, 
Phys. Rev. Lett. {\bf 70}, 2182 (1993)] with two important
modifications due to cell-to-cell diffusive coupling. First, this 
coupling breaks the degeneracy of an infinite-dimensional Hopf bifurcation
such that the most unstable mode of alternans corresponds to
the longest quantized wavelength of the ring.
Second, the Hopf frequency, which determines the velocity
of the node along the ring, depends both on the steepness of
CV-restitution and the strength of this coupling, with the net result that
quasiperiodic behavior can arise with a constant conduction
velocity. In both the paced geometries and the ring, the onset of 
alternans is different in tissue than for a paced isolated cell.
The implications of these results for 
alternans dynamics during two-dimensional reentry are
briefly discussed.
\end{abstract}


\pacs{PACS numbers: 87.19.Hh, 05.45.-a, 05.45.Gg, 89.75.-k}

\maketitle

\section{Introduction}

Beat-to-beat changes in the T-wave portion of the electrocardiogram (ECG)
that reflects the repolarization of the ventricles $-$a phenomenon known as 
T-wave alternans$-$\cite{Lewis}, 
can be a precursor to life-threatening
ventricular arrhythmias and sudden cardiac death 
\cite{precur}. T-wave alternans have been related to alternations of 
repolarization at the single cell level \cite{Pasetal}, thereby 
establishing a causal link between 
electrical alternans and the initiation of ventricular fibrillation. 
Repolarization alternans are characterized by a beat to beat oscillation 
of the
action potential duration (APD) at sufficiently short pacing interval 
\cite{Min}. Experiments in both two-dimensional \cite{Pasetal}
and linear strands \cite{Foxetal} of cardiac tissue,  
as well as ionic model simulations \cite{Foxetal,Quetal,Watetal}, have 
shown 
that the resulting sequence of long and short action potential durations 
can be either in phase
along the tissue (concordant alternans), or can split into extended 
regions
oscillating out of phase (discordant alternans). 
The latter case is of special importance since it can lead to 
conduction blocks \cite{Foxetal} and the initiation of ventricular 
fibrillation \cite{Pasetal}. Furthermore, 
alternans can provide a mechanism of wave
breaks that sustain ventricular fibrillation \cite{Kar94,FeCh02}. The dual role
of discordant alternans in the initiation and, potentially, 
the maintenance of fibrillation provides 
the motivation to develop a fundamental 
understanding of this phenomenon.

Experimental \cite{FraSim88} and 
theoretical studies \cite{Couetal,Vin} have shown that spatially modulated
patterns of alternans associated with quasiperiodic oscillations
of APD can arise naturally in a ring of tissue. 
The terminology of spatially ``discordant alternans'',
however, originates from the observation of such patterns in periodically
paced tissue \cite{Pasetal}. Their appearance in this context
was first attributed to preexisting spatial 
heterogeneities in the electro-physiological properties of the tissue 
\cite{Pasetal,Ro01}. Although heterogeneities may certainly be present 
\cite{LaGi96}, and can affect
the formation of discordant alternans \cite{Quetal}, numerical simulations 
of
ionic models in homogeneous tissue \cite{Foxetal,Quetal,Watetal,WeKa06} 
have 
shown that they are not necessary for their formation.  
Dynamical properties alone are able to produce the spatially heterogeneous
distributions of APD observed in the experiments.  

Pioneering studies by Nolasco and Dahlen \cite{NolDal}
and Guevara {\it et al.} \cite{Gueetal} first explained the
occurrence of alternans in a paced cell in terms of the restitution relation
\begin{equation}
{\rm APD}^{n+1}=f({\rm DI}^n),\label{res}
\end{equation}
between the action potential duration at the $n^{th}+1$ 
stimulus, ${\rm APD}^{n+1}$, and the so-called diastolic interval
${\rm DI}^n$. The latter is the time interval between 
the end of the previous ($n^{th}$) action potential and the
time of the $n^{th}+1$ stimulus during which the
tissue recovers its resting properties. The restitution
relationship describes only approximately the actual beat to beat
dynamics because of memory effects that have been modeled
phenomenologically through higher
dimensional maps \cite{ScCa04,Fenton,HaBa99,FoBo02,ToSc03}. 
Moreover, recent theoretical studies have highlighted the
role of intracellular calcium dynamics in the genesis of alternans 
\cite{ShWa03,ShSa05}. 
In the present paper, we develop the amplitude-equation
approach assuming that the simple
relationship given by Eq. (\ref{res}) holds. 

If the time interval between two consecutive stimuli is fixed,
$\tau={\rm APD}^n+{\rm DI}^n=const.$ for all $n$, then Eq. (\ref{res})  
yields the map ${\rm APD}^n=f(\tau-{\rm APD}^{n-1})$ that has
a period doubling instability if the 
slope $f'$ of the restitution curve, defined by 
${\rm APD}=f({\rm DI})$, evaluated at its
fixed point exceeds unity. As the slope of the restitution curve typically
increases when the pacing period is decreased, alternans may develop beyond a
critical value of the pacing rate. In extended tissue, the
velocity of the activation wavefront also depends on DI, so the
oscillations of APD induce 
oscillations in the local period of stimulation. This, in turn, acts as
a control mechanism that tends to create nodes in the spatial
distribution of APD oscillations, thus resulting in discordant
alternans. 

In a previous paper \cite{EcKa02a}, we sketched the derivation
of an amplitude equation that provides a unified framework to understand 
the
initiation, evolution, and, eventually, control of cardiac alternans
\cite{EcKa02b,Chretal06}. 
For one-dimensional paced tissue, the equation takes the form  
\begin{equation}
\tau \partial_t a=\sigma a - ga^3 -\frac{1}{\Lambda}\int^x_0 a(x')dx' 
-w \partial_x a +\xi^2 \partial^2_x a,\label{eq.aintro}
\end{equation}
where $a\simeq ({\rm APD}^{n+1}-{\rm APD}^{n})/2$ measures the amplitude of 
period doubling oscillations in
APD, so that nodes separating out-of-phase alternans region
correspond to $a=0$, 
$\tau$ is the pacing period, $\sigma$ and $g$ are coefficients
that can be obtained from the map (\ref{res}), $w$ and $\xi$ are lengthscales
that depend on the strength of diffusive coupling between cells,
and $\Lambda$ is determined by
the dependence of the conduction velocity (CV), denoted by $c$, on diastolic
interval. The CV-restitution curve $c({\rm DI})$ is also known as 
the dispersion curve in the excitable media literature.  It should be noted
that time can be treated as a continuous variable in this framework
because, even though the APD oscillates from beat to beat, the amplitude
$a$ of this oscillation defined above 
varies slowly over many beats close to the period doubling
bifurcation. It is this slow spatiotemporal evolution that is described
by Eq. (\ref{eq.aintro}). 

In this paper, we provide a detailed derivation of the amplitude-equation 
above and extend it to two-dimensional paced tissue.
In addition, we extend this approach to treat the important
case of one-dimensional reentry in a ring of tissue. 
For clarity of presentation, we will first derive the amplitude-equation 
for the ring geometry, and then show how to 
modify the boundary conditions on this equation to treat
paced tissue in one and two dimensions. 
Although not treated in this paper, the amplitude equation formalism can be extended
to higher dimensional maps (that include memory, or the effect of
intracellular calcium), provided that the primary bifurcation is period doubling.
 
The derivation of the amplitude-equation
follows the general amplitude
equation framework that has been widely used to study the evolution of
weakly nonlinear patterns in non-equilibrium systems \cite{CroHoh}.
In the present context of alternans, however, this 
derivation is made extremely difficult by 
the stiff nature of ionic models and the fact that the underlying
stationary state (with no alternans) corresponds to a train of
pulses. To surmount these difficulties, 
the derivation of the amplitude-equation proceeds in
two steps. First, spatially coupled maps, which 
describe the beat-to-beat dynamics of APD oscillations, including
the crucial effect of cell-to-cell coupling,
are derived from the underlying ionic models. Second, a weakly
nonlinear and multiscale analysis is used to derive amplitude
equations from these spatially coupled maps. The weakly nonlinear
nature of the expansion is valid close to the
onset of the period doubling bifurcation. The multiscale nature of
the expansion, which allows us to retain only the two lowest order spatial
derivative terms in Eq. (\ref{eq.aintro}), is itself justified by
the fact that the wavelength of spatial modulation of alternans is large
compared to the length scale $\sim \xi$ over which the action potential 
of a cell is influenced by neighboring cells through
diffusive coupling.  

The paper is organized as follows. 
In the next section, we introduce two ionic models and compute their 
action potential duration and conduction velocity
restitution curves. These curves are used to relate quantitatively
the ionic models and the amplitude equations. 
The models considered are the Noble model for
Purkinje fibers \cite{Nob} and a simple two-variable ionic model
that captures basic dynamical properties of 
the cardiac action potential. This model has the advantage that
it is simple enough to calculate analytically the restitution
curves as well as all the coefficients 
of the amplitude-equation (\ref{eq.aintro}). 

Section \ref{ampeq} is devoted to the derivation 
of the amplitude equation in the 
ring geometry.  This equation is then used to study
the stability of pulses. The results are compared
to previous analyses based on coupled maps 
\cite{Couetal,Vin}.  
Courtemanche {\em et al.} \cite{Couetal} showed that the
wavelength of spatial modulation of 
alternans is quantized in a ring geometry.
In the same language that has been used recently to describe discordant
alternans in paced tissue, this quantization condition  
corresponds, in the ring, to the existence of different modes of 
discordant
alternans with an odd number of moving nodes
separating out-of-phase regions of period doubling
oscillations. The amplitude equation sheds light on the
effect of cell-to-cell electrical coupling on the stability
of these quantized modes of alternans. In particular,
we find that this coupling lifts the degeneracy of an
infinite-dimensional Hopf bifurcation so that the mode
with one node is the most unstable, in qualitative
agreement with the prediction
of coupled maps that include a phenomenological 
description of cell-to-cell coupling
\cite{Vin}. Furthermore, we show that 
coupling can lead to quasiperiodicity even in 
the absence of CV-restitution. 
The effect of diffusive coupling on one-dimensional
alternans dynamics in tissue
has also been investigated in the context of a stability analysis
of FitzHugh-Nagumo type models \cite{CytKee02}. This analysis 
captures the fact that the onset of alternans in tissue can differ from
a single cell. However, it does not provide a 
general framework to make predictions relevant for more realistic
ionic models of cardiac excitation or experiments.  

We extend our theory to the paced case in section \ref{paced}. 
The main result is that discordant alternans results  
from a pattern-forming linear instability that produces
either standing, with fixed nodes, or traveling waves, with
moving nodes that 
separate out-of-phase alternans regions. Whereas in the ring
the wavelength of discordant alternans is determined by the
ring perimeter, the wavelength in the paced case is \emph{independent}
of tissue size. In the latter, 
spatial patterns form due to the competing effect
of CV-restitution, contained in the nonlocal term in Eq. 
(\ref{eq.aintro}), 
which tends to create steep spatial gradients of repolarization,
and diffusive cell-to-cell coupling contained in the spatial gradient
terms, which tends to smooth these gradients. As a result of this 
competition, 
the wavelength $\lambda$ of discordant alternans depends on the three 
fundamental
length scales $w$, $\xi$, and $\Lambda$ in Eq. (\ref{eq.aintro}), with
different scalings $\lambda\sim (w \Lambda)^{1/2}$ and 
$\lambda\sim (\xi^2\Lambda)^{1/3}$ for standing and traveling waves,
respectively. We 
also extend the model to two-dimensional paced tissue. 
This extension is straightforward under the assumption that the
propagation of the front is weakly perturbed by the oscillations in
APD. Finally, we discuss briefly nonlinear effects and conduction 
blocks.

Finally, in sections \ref{sec.disc} and \ref{sec.conc} we present the discussion and conclusions. 
Appendix A contains additional details on the derivation of the restitution 
curves for the two-variable ionic model. Appendix B, in turn, 
contains the details of the derivation of the integral kernel that couples 
spatially the iterative maps of the beat-to-beat dynamics. This derivation makes 
use of the Green's function for the diffusion equation to obtain an integral 
expression for the time course of the membrane voltage and hence the APD.

\section{Ionic models and restitution curves\label{ionmodels}}

Our formulation is based on the action potential
duration and conduction velocity restitution
properties. To obtain the restitution curves, we simulate
the ionic models in a one-dimensional strand of paced tissue.
We consider the standard cable equation
\be
\partial_t V = D\, \partial^2_x V 
-\left(I_{\rm ion}+I_{ext}\right)/C_m,
\label{cable}
\ee
with the membrane current $I_{\rm ion}$, time in units of millisecond 
(ms),
$D=2.5 \times 10^{-4}$ cm$^2$/ms, $C_m=12 \,\mu$F/cm$^2$.
The external current $I_{ext}$ models a sequence of stimuli applied   
at $x=0$ at a constant pacing interval $\tau$. We impose zero 
gradient boundary conditions on $V$ at the two ends of the cable. 

\begin{figure}[p]
\centerline{
\includegraphics[width=8cm]{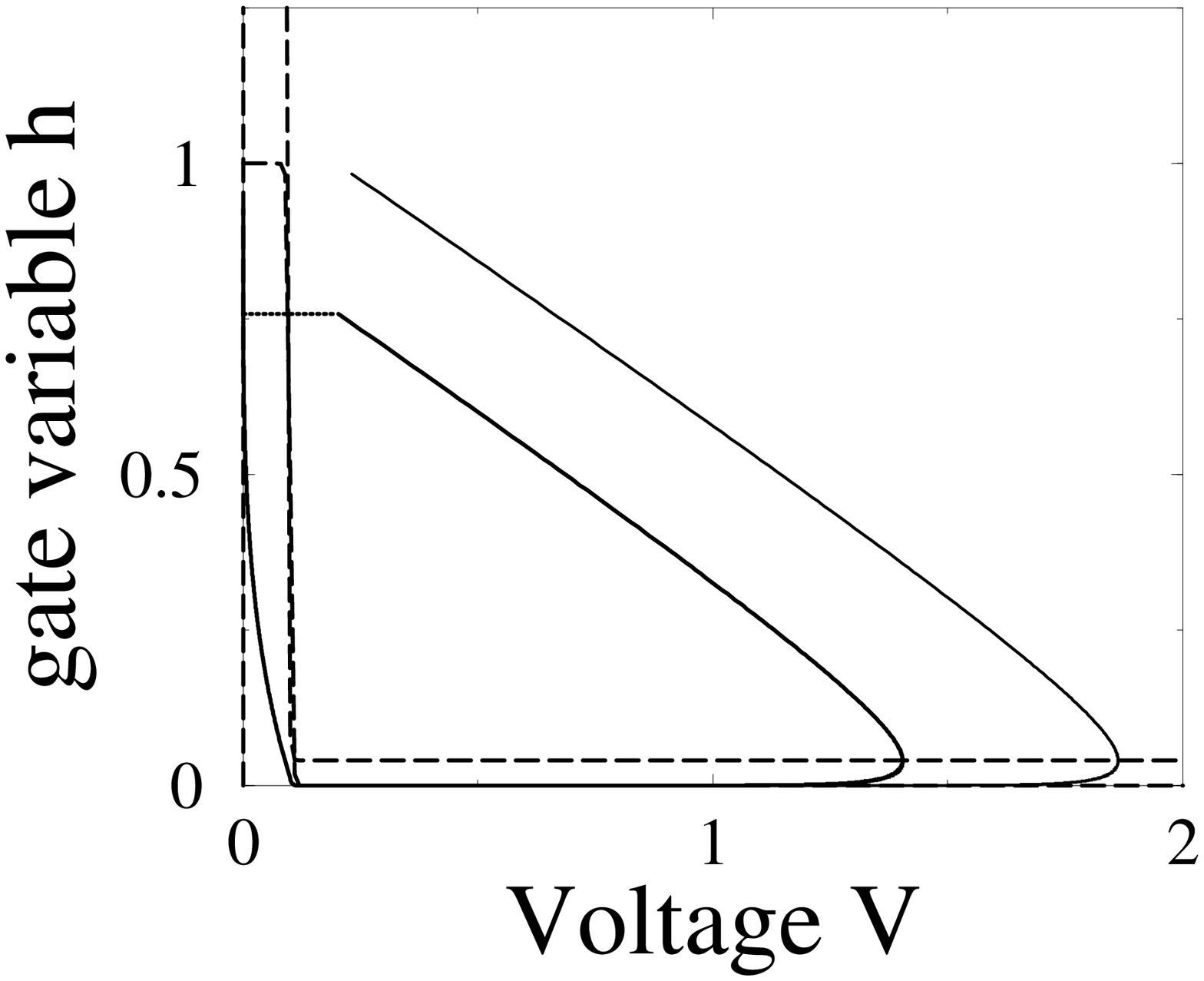}\hspace{0.5cm}
\includegraphics[width=8cm]{fig1b.eps}
}
\vspace{0.5cm}
\caption{a) Trajectories in the phase plane ($V$, $h$) for the 
two-variable
model [Eqs. (\ref{eq.I2var})-(\ref{eq.h2var})], obtained by simulating 
$\dot{V}=-I_{ion}/C_m$ with an activation interval $\tau=400~ms$ and
the parameters given in the text. 
The dashed lines denote
the nullclines $h=0$ and $V=0$. b) Time evolution
of the voltage $V$ (solid line) and the gate variable $h$ (dashed line) 
during such trajectory. 
\label{fig.ph-plane}}
\end{figure}

In the rest of the paper, we will use two models for the ionic currents.
The first is the original 1962 
Noble model for Purkinje fibers \cite{Nob}. The second is a simplified 
two-variable model with a triangular shaped action potential. This model
is more tractable analytically for the purpose of deriving spatially coupled maps 
of the beat-to-beat dynamics and the length scales $w$ and $\xi$ of the amplitude 
equation. For higher order ionic models like the Noble model, these coefficients 
need to be obtained numerically following a procedure detailed in Sec. IV.A.  The 
two-variable 
model is defined by the 
total membrane current \cite{EcKa02a}
\begin{equation}
\frac{I_{ion}}{C_m}=\frac{1}{\tau_0}\left [S+(1-S)V/V_c\right ]
  +\frac{1}{\tau_a} h \,S. 
\label{eq.I2var}
\end{equation}
This current is the sum of a slow, time-independent, outward current
(similar to the potassium current in more realistic ionic models), which
repolarizes the cell on a slow time scale $\tau_0$, and a fast inward 
current that is a simplified version of the sodium current. The latter 
depolarizes the cell in the fast time $\tau_a \ll \tau_0$. The 
inactivation of the fast inward current 
is regulated by the gate variable $h$, which evolves according to 
\begin{equation}  
\frac{d h}{dt}=\frac{1-S-h}{\tau_{-}(1-S)+S\tau_{+}},
\label{eq.h2var}
\end{equation}
where $\tau_{+}$ and $\tau_{-}$ control the time scales of
inactivation and recovery from inactivation of this current.
In this model the transmembrane voltage $V$ is dimensionless, and 
$S\equiv [1+\tanh((V-V_c)/\epsilon)]/2$ is a sigmoidal function. We consider
the values of
the parameters shown in Table \ref{param2var}.
Note that the product of the $h$ and $j$
gates in standard formulations of the sodium current 
is represented, in the present model, 
by a single gate $h$. Thus choosing $\tau_{-}$ larger than
$\tau_{+}$ produces the same effect of a  
$j$ gate that controls a slower recovery 
from inactivation in comparison to inactivation controlled by
the $h$ gate. Fig. \ref{fig.ph-plane} illustrates the triangular shaped
action potentials obtained with this model for
stimuli spaced by 400 ms. 

\begin{table}
\begin{tabular}{|l|l|l|}
\hline
$V_c=0.1$ & $\epsilon=0.005$ & $\tau_0=150$ ms  \\ \hline
$\tau_a=6$ ms & $\tau_{-}=60$ ms & $\tau_{+}=12$ ms \\ \hline 
\end{tabular}
\caption{Parameters used in the simulations of the two-variable model
  [Eqs. (\ref{eq.I2var})-(\ref{eq.h2var})]}
\label{param2var} 
\end{table}

\begin{figure}[p]
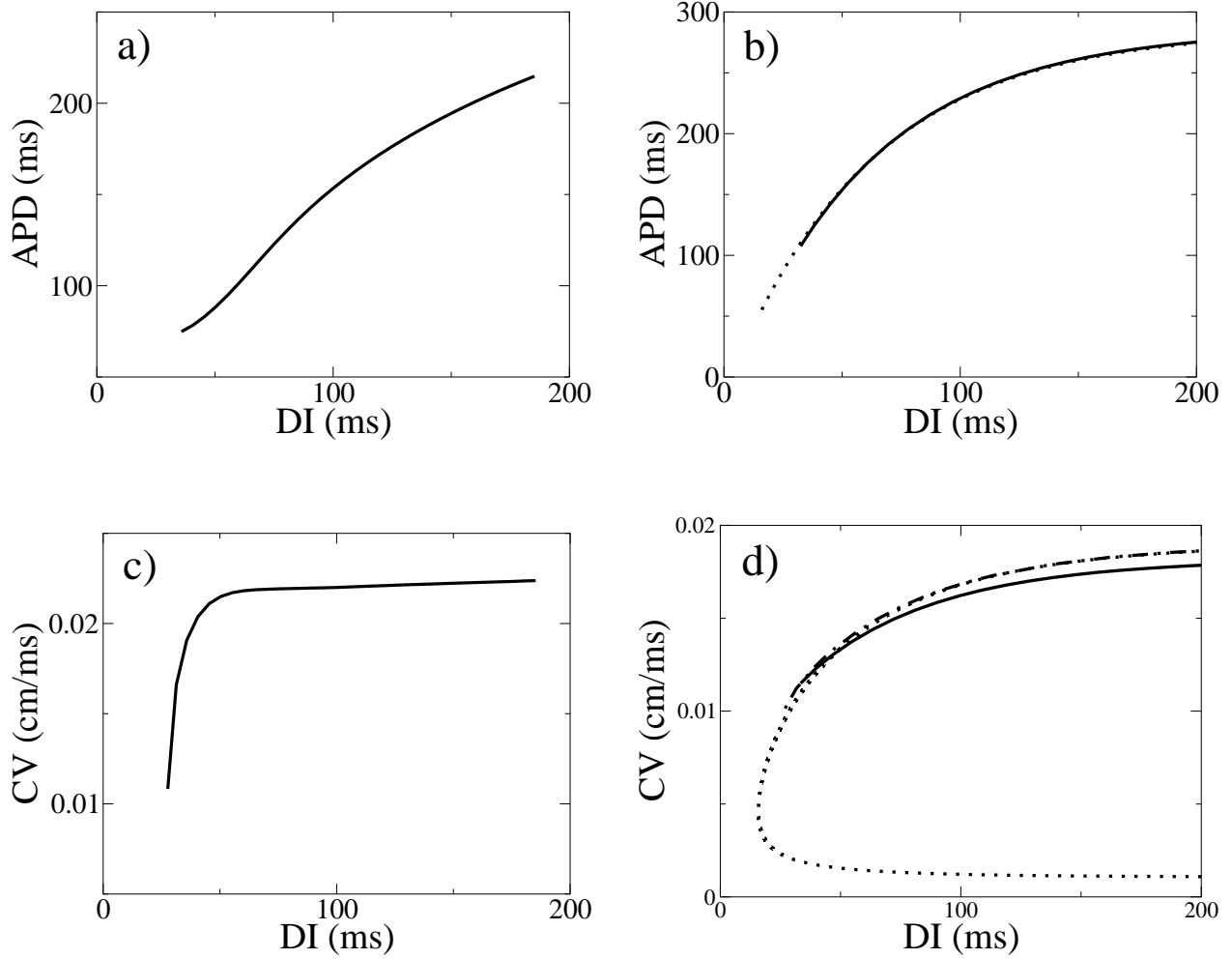

\centerline{
\includegraphics[width=8cm]{fig2a.eps}\hspace{0.5cm}
\includegraphics[width=8cm]{fig2b.eps} 
}
\vspace{1cm}
\centerline{
\includegraphics[width=8cm,clip=true]{fig2c.eps}\hspace{0.5cm}
\includegraphics[width=8cm,clip=true]{fig2d.eps}
}
\caption{APD- and CV-restitution curves corresponding to the Noble [a) and c)]
and two-variable model [b) and d)]. The solid lines correspond to the
restitution curves obtained numerically using a S1-S2 protocol. In b) and d)
the dotted lines correspond 
to the theoretical curves given by Eqs. (\ref{ec.restAPDcable}) and
(\ref{restCVcable}). The dotted-dashed line in d) corresponds to simulations
with a 
finer grid dx=0.001, and dt=0.001, which agree very well 
with the theoretical predictions, and indicate that, for the usual parameters
in the simulations, there is an error of
about a 4\% in the determination of the maximum conduction velocity.}
\label{fig.rest}
\end{figure}

We have simulated Eq. (\ref{cable}) using the forward Euler method, with
a three-point finite difference representation of the one-dimensional 
Laplacian. We take a mesh size $dx=0.01$ cm, and a time step $dt=0.02$ ms 
in the case of the two-variable 
model, and $dx=0.01$ cm, $dt=0.05$ ms, for the Noble model. The thresholds 
of the transmembrane voltage to define the APD and DI in the two-variable
model and the Noble model are $V=V_c=0.1$ and 
$V=-40$ mV, respectively. The restitution and dispersion curves are 
computed by pacing Eq. (\ref{cable}) in a short cable by an S1-S2 protocol,
i.e., the cable is paced at a large period for several beats (sufficient to
achieve steady state APD), and then an extrastimulus (S2) is delivered at
progressively shorter coupling intervals to vary DI  
(see Fig. \ref{fig.rest}). The stimulus is
typically applied over ten grid points. 

In the limit in which the sigmoidal function $S\equiv
(1+\tanh((V-V_c)/\epsilon))/2$ becomes a step function ($\epsilon
\;\rightarrow \;0$), it is possible
to calculate both the APD- and
CV-restitution curves of the two-variable model analytically (see Appendix
\ref{ap2var}), provided that the ${\rm APD}$ is much larger
than the time scale of inactivation of the fast inward sodium
current, or ${\rm APD} \gg \tau_{+}$, and that $D/c$ is much smaller
than the wavelength of the action potential, or $D/c \ll c {\rm APD}$.
Both conditions are satisfied for physiologically relevant conditions. 
In this limit, the restitution 
curve is found to be well approximated by the form
\be
{\rm APD} \simeq \frac{\tau_{+}\tau_0}{\tau_a}[1-\exp(-{\rm DI}/\tau_{-})]-V_c 
\tau_0.\label{eq.2var_APD}
\ee
Accordingly, the onset of alternans is given by
\be
\frac{{\rm d}\,{\rm APD}}{{\rm d}\,{\rm DI}}\simeq\frac{\tau_0 \tau_{+}}{\tau_a\tau_{-}}
\exp(-{\rm DI}/\tau_{-})=1,
\ee  
or
${\rm DI}_c\simeq\tau_{-}\log[(\tau_0\tau_{+})/(\tau_a\tau_{-})]$, which yields 
${\rm DI}_c\simeq 96.6$ ms and ${\rm APD}_c\simeq 225$ ms 
for the two-variable model parameters of Table \ref{param2var}.

In the same limit, the CV-restitution curve can be calculated considering a 
traveling pulse
$V(x,t)=V(x-ct)$, and matching the expressions obtained for $V>V_c$ 
and $V<V_c$, in the wavefront and the waveback (appendix \ref{ap2var}). This
results  
in an implicit equation for $c$ (cf. Eq. (\ref{restCVcable})), that agrees
well with the numerical results (see Fig. \ref{fig.rest}d).

\section{Ring geometry\label{ampeq}}

The stability of pulses circulating in a ring of tissue was
considered previously by Courtemanche {\it et al.}
\cite{Couetal} using generic restitution properties of the system
as given by Eq. (\ref{res}). This stability analysis was based on the
idea to unravel the ring into an infinite line (then $x\in {\cal R}$, 
instead of
$x\in {\cal R}\; {\rm mod}\; L$). 
In this manner, the
values of a given variable at a previous passage of the pulse through a 
given 
point $x$ in the ring, and therefore, at a previous beat, can be
identified with the value of that variable 
at the point $x-L$ (i.e. ${\rm APD}^{n}(x)\equiv {\rm APD}(x+nL)$). 
This allows to drop the dependence of the different variables on the beat 
number $n$, and reformulate the maps as delay differential equations. 
Using this approach, Courtemanche {\it et al.} 
found that, when the slope of the
restitution curve is greater than one, the pulses become unstable
towards modes of propagation where the values of the APD and
conduction velocity vary in space around a mean value. Because of the
periodic boundary conditions, the wavelength of these
modes is quantized. Without CV-restitution, the 
ratio of the wavelength to twice the ring perimeter 
is $1/(2n+1)$ where $n\ge 0$ in an integer. 
With CV-restitution at onset
$c'({\rm DI}_c)\equiv {\rm d}\,c/{\rm d}\,{\rm DI}({\rm DI}_c)\neq 0$, this ratio becomes
irrational and the nodes travel, giving rise to 
quasiperiodic 
motion at a given point. 

In the analysis of \cite{Couetal} all the modes
were found to bifurcate simultaneously 
at the same ring perimeter, in an infinitely degenerate
Hopf bifurcation. This degeneracy was shown to be broken \cite{Vin} by
the effect of diffusive coupling, which effectively modifies the
restitution relation (\ref{res}) and makes the mode with the longest
wavelength (with a single node) bifurcate first. In \cite{Vin}, the effect of 
diffusive coupling was considered phenomenologically, assuming a coupling of 
the APD in neighboring cells 
given by a simple gaussian kernel. Here, we derive the form of this
kernel from an analysis of the two-variable model.
The most interesting finding is that this kernel is generally 
asymmetrical due to the fact that parity ($x\rightarrow -x$) symmetry
is broken by the propagation direction of the action potential;
it only reduces to a symmetrical gaussian kernel in the limit
of infinite conduction velocity. 
In effect, the way a given 
cell is influenced by its left and right neighbors is different 
because these cells are activated at different
times by the propagating wavefront. The most interesting consequence of
this asymmetry is that it can produce quasiperiodic 
motion even in the absence of CV-restitution, and modifies the
frequency of quasiperiodic oscillations in the presence of 
CV-restitution.      

To study the stability of the pulses we will use an approach slightly 
different from that in \cite{Couetal}. We will convert the ring into a 
linear cable of length $L$ with origin at $x=0$, and index with $n$
the number of times the pulse has traveled around the ring. 
Accordingly, the variables that characterize the beat-to-beat dynamics
depend both on the beat number $n$ and on the space
variable $0\le x \le L$; the diastolic interval is written
as ${\rm DI}^n(x)$ and the ring geometry imposes the periodic boundary
condition ${\rm DI}^n(x-L)={\rm DI}^{n-1}(x)$.   
Using this parametrization, the ring dynamics is governed by
the equations
\begin{eqnarray}
T^n(x)&=& \int_{x-L}^x \frac{d\,x'}{c[{\rm DI}^n(x')]}=\int_x^L 
\frac{d\,x'}{c[{\rm DI}^{n-1}(x')]}+\int_0^x
\frac{d\,x'}{c[{\rm DI}^n(x')]} \label{disp},\\ 
{\rm APD}^{n+1}(x)&=&\int^{\infty}_{-\infty} G(x'-x) f[{\rm DI}^{n}(x')]\,dx',
\label{eq.kernel}
\end{eqnarray}
where $T^n(x)$ is the time interval between two consequent activations
at the same point.
The first equation is just a kinematic equation stating that the
period of stimulation at a given point is given by the time it takes a
pulse to complete a revolution. The second equation reflects the 
restitution properties of the system, 
where we have included an asymmetrical kernel $G(x'-x)$ that appears
because of the diffusive coupling between neighboring cells. 
In general, the derivation of the kernel $G(x'-x)$ is very 
difficult and we only derive it explicitly in Appendix \ref{appA} for our 
two-variable model. The calculation of the kernel involves
the inversion of a nonlocal, implicit equation for the APD. Even though
we cannot derive this kernel for a general higher order 
ionic model, this is not a serious limitation. Only the length scales
$w$ and $\xi$ in Eq. (\ref{eq.aintro}), but not the general form of the
amplitude-equation,  
depend on a precise knowledge of this kernel. Moreover,  
$w$ and $\xi$ can be calculated numerically for a general 
higher order ionic model using the 
procedure detailed in Sec. IV.A.   

\subsection{Linear stability revisited}

Let us thus start reviewing the linear stability analysis of
Eqs. (\ref{disp}) and (\ref{eq.kernel}).
Taking into account that $T^n(x)={\rm APD}^n(x)+{\rm DI}^n(x)$ one can write 
Eqs. (\ref{disp}), (\ref{eq.kernel}) as a single equation for 
${\rm DI}^{n+1}(x)$:
\begin{equation}
{\rm DI}^{n+1}(x)=\int_{x-L}^x \frac{d\,x'}{c[{\rm DI}^{n+1}(x')]}-
\int^{\infty}_{-\infty} G(x'-x) f[{\rm DI}^{n}(x')]\,dx'.
\label{eq.DIring}
\end{equation}
Then, a stationary solution satisfies
${\rm DI}^*=\tau-f({\rm DI}^*)$,
where $\tau=L/c({\rm DI}^*)$ is the period of propagation of the pulse, and
we choose the kernel $G(x'-x)$ to be normalized, so
\be
\int^{\infty}_{-\infty} G(y)dy=1.
\ee
We will also assume that the kernel has compact support and that it decays
fast enough.

Perturbing around the stationary solution in the form ${\rm DI}^n(x)={\rm DI}^*+
\alpha^n e^{ikx} \delta D $, then the stability of the steady state solution
is dictated by the value of $\alpha$. If $|\alpha|>1$ an initially small
perturbation will grow. Alternans occurs when $\alpha=-1$, and then the
resulting 
instability gives a beat to beat change in the amplitude of APD. Inserting
the former expansion in Eq. (\ref{eq.DIring}) and linearizing, we obtain that the  
following equation must be satisfied
\begin{equation}
\alpha =\frac{i\alpha }{2\Lambda k}(1-e^{-ikL}) -f'\tilde{G}(k),
\label{carac.ring1}
\end{equation}
where we have defined the Fourier transform of the kernel
\begin{equation}
\tilde{G}(k)=\int_{-\infty}^{\infty}G(y)e^{iky}dy.\label{eq.G(k)} 
\end{equation}
Here $\Lambda\equiv  c^2/(2\,c')$, defined with $c$ and $c'\equiv {\rm
  d}c/{\rm dDI}$ 
evaluated at the bifurcation point, is a characteristic length scale associated 
with CV-restitution.

Furthermore, from the condition ${\rm DI}^{n+1}(x-L)={\rm DI}^{n}(x)$ we obtain
the quantization condition $e^{ikL}=\alpha$, so Eq. (\ref{carac.ring1}) can be
rewritten as 
\begin{equation}
\alpha (1-\frac{i}{2\Lambda k})=-f'\tilde{G}(k) -\frac{i}{2\Lambda k}.
\label{carac.ring2}
\end{equation}
The onset of the instability is given by $|\alpha|=1$ (implying that
the bifurcating mode will have a real wavenumber $k$). This yields   
\begin{equation}
f'^2 |\tilde{G}(k)|^2+\frac{f'}{\Lambda k}{\cal I}m
[\tilde{G}(k)]=1.\label{ins.ring}
\end{equation}
The quadratic equation can be solved to obtain the value of $f'$:
\begin{equation}
f'=\frac{1}{2|\tilde{G}(k)|^2} \left ( -\frac{{\cal I}m[\tilde{G}(k)]}{\Lambda
    k} \pm \sqrt{({\cal I}m [\tilde{G}(k)])^2/(\Lambda k)^2+4|\tilde{G}(k)|^2}
    \right ). 
\label{eq.fprimgen}
\end{equation}
The imaginary part of the kernel appears because of asymmetrical coupling.
Due to the finite velocity of propagation of the pulses, different points are
excited at different times, and the electrotonic coupling of a cell with its
left- and right-neighbors will differ. We can assume that this is a small
effect, and then:
\begin{equation}
f' \simeq \frac{1}{|\tilde{G}(k)|}- \frac{{\cal I}m [\tilde{G}(k)]}{2\Lambda k
  |\tilde{G}(k)|^2},
\label{eq.fprimpar}
\end{equation} 
where we have taken the plus sign in Eq. (\ref{eq.fprimgen}), corresponding to
alternans 
(period doubling). The minus sign would correspond to a steady
instability, not studied in this paper.

Since the kernel decays on a spatial scale $\xi$ much shorter
than the wavelength of the unstable modes of interest ($\sim L$), 
the exponential in Eq. (\ref{eq.G(k)}) can be expanded in the form
$e^{iky} \simeq 1+iky - (ky)^2/2 + \cdots$. It then follows that
\begin{equation}
\tilde{G}(k)\simeq 1 -iw k - \xi^2 k^2, \label{kernel.four}
\end{equation}
where we have defined the coefficients
\begin{equation}
w=-\int_{-\infty}^{\infty}G(y)ydy,\;\;\;\;
\xi^2=\frac{1}{2}\int_{-\infty}^{\infty}G(y)y^2dy.\label{def.coeffs}
\end{equation}
For an arbitrarily complex ionic model, 
the form of the kernel $G(y)$ 
cannot be calculated explicitly, and the coefficients $w$ and $\xi^2$
must be obtained from the numerical simulation of the ionic model
as described in Sec. IV.A. For
the two-variable model 
with a constant repolarization rate, however, they can be calculated 
explicitly (Appendix \ref{appA}), giving 
\begin{eqnarray}
w&=&2D/c,\label{w2v} \\
\xi &=&(D\times  {\rm APD}_c)^{1/2}. \label{xi2v}
\end{eqnarray}
Eq. (\ref{xi2v}) has the simple physical interpretation that the
transmembrane potential $V$
diffuses a length $\sim \xi$ in the time interval of one 
APD. Therefore, the repolarization of a given cell
is influenced by other cells within a length $\sim \xi$ of cable.
The imaginary part of $\tilde{G}(k)$ appears because of the asymmetry 
induced by the propagation of the pulse. This asymmetry vanishes 
in the limit $c\rightarrow \infty$, where all cells are activated 
simultaneously, consistent with Eq. (\ref{w2v}). For typical values of the
parameters in ventricular tissue ($D \sim 10^{-4}-10^{-3}\, \rm{cm}^2/\rm{ms}$, 
$c \sim
10^{-2}-10^{-1}$ cm/ms, ${\rm APD}_c \sim 100-200$ ms), the length $\xi$ is of the
order of millimeter ($\xi \sim 10^{-1}$ cm), while $w$ an order of magnitude 
smaller ($w \sim 10^{-2}$ cm). In fact, $w/\xi \sim \xi /(c \times {\rm APD}_c)$
is the ratio between the diffusive coupling length $\xi$ and the wavelength 
of the pulse, which is typically an order of magnitude larger.

Introducing expansion (\ref{kernel.four}) into Eq. (\ref{eq.fprimpar}), to
lowest order in $k\xi$, we obtain:
\begin{equation}
f'=1-\frac{w}{2\Lambda}+\xi^2 k^2.\label{eq.f'}
\end{equation}
Intercellular coupling shifts the onset of
instability, which occurs for a value of the slope of restitution
$f'\neq 1$. The slope at onset of alternans will be larger or smaller than
one, depending on the relative importance of the different lengthscales of the
system. The first mode to bifurcate is the one with largest
wavelength, as previously noted by Vinet \cite{Vin}.  

In what follows, in addition to assuming that 
$wk \ll 1$ and $\xi k \ll 1$, we will also assume that the slope
of the CV-restitution curve at the bifurcation point is small  
so that the length scale $\Lambda$ is much larger that the typical wavelength of
alternans modulation, or $\Lambda k \gg 1$. All three conditions turn out
to be reasonably well satisfied for the parameters of the Noble model. 
The last condition $\Lambda k \gg 1$, however, is not fulfilled in
the two-variable model because CV-restitution is not weak. In the next section,
we will show that it is also possible to treat analytically the case where
$\Lambda k$ is of order unity.  

Substituting Eq. (\ref{eq.f'}) into Eq. (\ref{carac.ring2}), to lowest
order we obtain 
\begin{equation}
\alpha\simeq -1+iwk-\frac{i}{\Lambda k}.\label{alfring}
\end{equation}
Both dispersion and asymmetrical
coupling result in an imaginary part for $\alpha$, that corresponds to
quasiperiodic motion. 

If $\alpha=-1$, the instability is strictly 
period doubling. In this case, the condition $e^{ikL}=\alpha$ implies  
$k=(2n+1)\pi/L$, with $n=0,1,2,\dots$. To obtain the wavenumber $k$
corresponding to the value of $\alpha$ given in Eq. (\ref{alfring}), we assume
$k=(2n+1)\pi/L+q$, where $q$ is a small correction $q \ll 1/L$. In this case the
condition $e^{ikL}=\alpha$ results into 
\begin{equation}
e^{ikL}= (-1) e^{iqL} \simeq (-1) (1+iqL +\cdots) \simeq  -1+iw(2n+1)\frac{\pi}{L}
-\frac{iL}{\Lambda(2n+1)\pi} + \cdots. 
\end{equation}
Then, to first order in the correction, the wavenumber is  
\begin{equation}
k=(2n+1)\frac{\pi}{L}+\frac{1}{\Lambda(2n+1)\pi}-w(2n+1)\frac{\pi}{L^2}.
\end{equation}
As $k \sim L^{-1}$, these expressions will be valid if $w,\xi \ll L$, and
$\Lambda \gg L$.  

The main novel conclusion is that 
besides the correction to the wavenumber due to CV-restitution derived by
Courtemanche {\it et al.} \cite{Couetal}, there is another correction due to 
asymmetrical coupling. Therefore quasiperiodic motion can occur even with a 
constant CV. Furthermore, these two effects can balance each other if
\be
w=\frac{L^2}{\Lambda(2n+1)^2 \pi^2},\label{ring.period}
\ee
in which case the motion becomes strictly periodic even 
with a finite amount of CV-restitution. In general, these two effects will
not exactly balance each other so that the motion will be quasiperiodic with a 
frequency determined by both the slope of the CV-restitution curve and the 
asymmetrical coupling strength $\sim w$.

\subsection{Derivation of the amplitude equations}

Starting from the maps (\ref{disp}) and (\ref{eq.kernel}), it is
possible to derive  
equations for the oscillations in period and action potential duration. 
To that end, we will consider the second iteration of the map 
(\ref{eq.kernel}), and expand it for small values of the amplitude of 
oscillation.
As the change in the value of the APD every two beats is small, the beat
number can be treated as a continuous variable. Also, the dispersion
relation for the conduction velocity $c=c({\rm DI})$ can be expanded for
small oscillations of DI, obtaining from Eq. (\ref{disp}) the 
corresponding change in the
local activation interval, which depends nonlocally on the oscillations
of APD. The only non-trivial point in 
the expansion is the effect of the non-local electrotonic coupling in Eq. 
(\ref{eq.kernel}). Since the 
kernel decays on a scale $\sim \xi$ shorter than the wavelength of
modulation of alternans, it can expanded to obtain a local 
relation between the change in APD and its local
first and second spatial derivatives. 

Close to the onset of oscillations we can write
\begin{eqnarray}
{\rm APD}^n(x)&\approx &{\rm APD}_c\,+\,\delta A\,+\,(-1)^n\, a(x,t), \label{aa}\\
T^n(x)&\approx &\tau_c\,-\delta \tau\,+\,(-1)^n\, b(x,t),\label{bb}
\end{eqnarray}
where ${\rm APD}_c$ and $\tau_c$ are the APD and the period of stimulation 
evaluated at the bifurcation point of the single-cell map 
($f'=1$), and we split the perturbations into steady and oscillating parts. Now 
the basic 
pacing period will be the traveling
time of a pulse around the ring, which in the absence of oscillations is
given by $\tau=L/c$, and $\delta\tau\equiv \tau_c-\tau\ll \tau_c$. 
Close to the bifurcation point, the steady correction to the APD is, to
first order, $\delta A= - \delta \tau/2$. For the oscillating part, 
since the beat to beat oscillations are taken into account with the
terms $(-1)^n$, the amplitude of the deviations from the critical values, 
$a(x,t)$ and $b(x,t)$, vary slowly from beat to beat. We can therefore
assume that $a$ and $b$ depend on a continuous time, defined through 
$n\equiv t/\tau$. 

Let us first discuss what are the boundary conditions satisfied by 
$a(x,t)$ and $b(x,t)$. The transmembrane voltage  
obeys periodic boundary conditions $V(L)=V(0)$,
but, by definition, after a revolution of the pulse along the ring, 
the system goes into the next beat.
Therefore, the values of the period and APD at $x=L$ (that is, right
before 
the end of the revolution) must equal those at $x=0$, 
at the next beat (i.e. at the beginning of the next revolution). The same
is true for the gradients. Then
\bea
&&T^n(L)=T^{n+1}(0),\;\;\partial_x T^n(L)=\partial_x T^{n+1}(0),\\
&&{\rm APD}^n(L)={\rm APD}^{n+1}(0),\;\;\partial_x {\rm APD}^n(L)=\partial_x
{\rm APD}^{n+1}(0).
\eea
Using Eqs. (\ref{aa}), (\ref{bb}) it is easy to see that, in terms of the 
oscillations in APD and period, the former boundary conditions become
\bea
&&a(L)=-a(0),\;\;\partial_x a(L)=-\partial_x a(0), \label{bca} \\
&&b(L)=-b(0),\;\;\partial_x b(L)=-\partial_x b(0). \label{bcb}
\eea
These boundary conditions imply that the pattern must have an odd number of
nodes. Namely, letting $a(x)\sim e^{ikx}$, one obtains the quantization
condition
\be
k=(2n+1)\frac{\pi}{L}, \;\;n=0,1,2,\dots,
\label{eq.cabk}
\ee
that obtained in \cite{Couetal} in the limit of zero
CV-restitution slope. 
As we shall see, the corrections to the 
wavelength come from a phase shift in the 
slow scale associated with the quasiperiodic motion. 

The equation for the oscillations in period is easy to obtain. First, we 
can
write Eq. (\ref{disp}) in differential form
\be
\frac{dT^n(x)}{dx}=\frac{1}{c[{\rm DI}^n(x)]}-\frac{1}{c[{\rm DI}^n(x-L)]}=\frac{1}{c[
{\rm DI}^n(x)]}-\frac{1}{c[{\rm DI}^{n-1}(x)]},
\label{eq.dispdiff}
\ee
where we take into account that ${\rm DI}^n(x-L)={\rm DI}^{n-1}(x)$. Then,
substituting expansions (\ref{aa}) and (\ref{bb}) into the former 
expression, with 
${\rm DI}^n(x)=T^n(x)-{\rm APD}^n(x)$, we obtain, at linear order
\be 
\frac{db}{dx}= \frac{1}{\Lambda} [a(x)-b(x)].\label{beqa}
\ee 
To obtain an expression for $b(x)$ 
we have to solve Eq. (\ref{beqa}), subject to the
boundary condition (\ref{bcb}). This results into:
\begin{equation}
b(x)=\frac{1}{\Lambda}\int^x_0 e^{(x'-x)/\Lambda} a(x') dx' -
\frac{e^{-L/\Lambda}}{\Lambda(1+e^{-L/\Lambda})}\int^L_0 e^{(x'-x)/\Lambda} a(x') dx'.
\label{beqaint}
\end{equation}
In order to be able to get an analytical 
expression for the shift in wavelength, we will take the limit of 
small dispersion ($L/\Lambda \ll 1$), that will also allow us to compare 
with the results in \cite{Couetal}. In this limit, to first order, the 
exponentials in
Eq. (\ref{beqaint}) can be neglected (equivalent to neglecting the term 
proportional
to $b(x)$ 
in the right hand side of Eq. (\ref{beqa})). Then, Eq. (\ref{beqaint}) becomes
\be
b(x)=\frac{1}{\Lambda}\int^x_0 a(x') dx' -
\frac{1}{2\Lambda}\int^L_0 a(x') dx'.
\label{eq.dba}
\ee
The above equation works well for the Noble model.
For the two-variable model, however, it is 
not strictly valid since $L_c/\Lambda > 1$. In this case,
as we discuss below, the full equation (\ref{beqaint}) 
should be used to obtain a good agreement with the
simulations of the cable equation (\ref{cable}). For clarity
of exposition, we use Eq. (\ref{eq.dba}) in what follows
and state the result later for the full equation (\ref{beqaint}) in
the context of a quantitative comparison of the two-variable
model and cable equation simulations.  

Next, in order to derive an evolution equation for 
the amplitude $a(x,t)$, we notice that, after two consecutive
beats, the APD becomes
\begin{equation}
{\rm APD}^{n+2}={\rm APD}_c+(-1)^{n+2}a(x,t+2\tau).
\end{equation}
Assuming that the APD varies slowly every two beats 
(which is the case close to the period doubling bifurcation), we
can expand $a(x,t+2\tau)\simeq a(x,t)+2\tau \partial a/\partial t$, so
\begin{equation}
{\rm APD}^{n+2}={\rm APD}^{n}+(-1)^n 2\tau \frac{\partial a}{\partial t}.\label{atime}
\end{equation} 
But, expanding Eq. (\ref{eq.kernel}) 
\be
{\rm APD}^{n+1}(x)=\int^{\infty}_{-\infty} G(y) f[{\rm DI}^{n}(y+x)]\,dy\simeq 
f[{\rm DI}^n(x)]-wf'\partial_x {\rm DI}^n(x)+\xi^2f'\partial^2_x {\rm DI}^n(x),
\label{expkernel}
\ee
we can also write ${\rm APD}^{n+2}$ as
\bea
{\rm APD}^{n+2}&=&f[T^{n+1}-f(T^n-{\rm APD}^n)+wf'\partial_x {\rm DI}^n 
-\xi^2f' \partial^2_x {\rm DI}^n]\nonumber\\
&&-wf'\partial_x (T^{n+1}-{\rm APD}^{n+1})+\xi^2f' 
\partial^2_x(T^{n+1}-{\rm APD}^{n+1}), 
\eea
where we have used the definitions in Eq. (\ref{def.coeffs}) for $w$
and $\xi^2$.
Then, using Eqs. (\ref{aa})-(\ref{bb}), expanding to first order in $w$ 
and
$\xi^2$, and taking into account that $f'=1$, since we are expanding 
around the
period doubling bifurcation, we have
\bea
{\rm APD}^{n+2}&=&f[T^{n+1}-f(T^n-{\rm APD}^n)]+w\partial_x {\rm DI}^n-\xi^2
\partial^2_x {\rm DI}^n \nonumber\\
&&-w\partial_x[T^{n+1}-f(T^n-{\rm
  APD}^n)]+\xi^2\partial^2_x[T^{n+1}-f(T^n-{\rm APD}^n)]
\nonumber\\
&=&f[T^{n+1}-f(T^n-{\rm APD}^n)]+(-1)^n \left [-w (2\partial_x a -3\partial_x 
b)
+\xi^2(2\partial^2_x a -3\partial^2_x b)\right ].\label{adevs}
\eea
In the following we will neglect the term $\partial_x b$ compared to
$\partial_x a$, since from (\ref{eq.dba}) we have that $b \sim a/(\Lambda k) \ll
a$, in the limit of small dispersion (i.e. weak CV-restitution). Strictly,
the amplitude equation we derive is asymptotically valid close to the
bifurcation point if the following
scaling relations are satisfied: $a/{\rm APD}_c \sim \epsilon^{1/2}$, $b/a \sim
1/(\Lambda k)\sim \epsilon$, $\xi k \sim \epsilon^{1/2}$, and $wk
\sim \epsilon$, where $\epsilon\sim \delta \tau/\tau_c$ is a dimensionless 
measure of the distance from the bifurcation point.

Equating now Eqs. (\ref{atime}) and (\ref{adevs}), and expanding the second 
iteration of the 
map, we obtain the final expression for the evolution of the oscillations 
of APD, in the limit considered,
\be
\tau \partial_t a =\sigma a -w\partial_x a +\xi^2 \partial^2_x a -g a^3 - 
b  ,\label{aeqb}\\
\ee
where $\sigma\equiv f''(\tau - \tau_c)/2$ is of order $\epsilon$, 
$g\equiv f''^2/4-f'''/6$, and all the derivatives are evaluated at the 
bifurcation point. These 
coefficients can be calculated from the curves in Fig. \ref{fig.rest}.
We find the bifurcation in the Noble model to 
be slightly subcritical, so Eq. (\ref{aeqb}) must be expanded to fifth 
order 
in this case \cite{coeffs}. For simplicity of exposition, in the following
we will focus on the supercritical case. When dispersion is not small,
one should keep higher order terms in Eq. (\ref{aeqb}) involving
the oscillation in period $b$. In that case, direct simulation
of the original coupled maps (\ref{disp})-(\ref{eq.kernel}) is
probably more appropriate, if the goal is to achieve good
quantitative agreement with ionic model simulations. 

Substituting the expression for $b(x)$ into Eq. (\ref{aeqb}), 
we obtain the final amplitude equation in the ring geometry
\be
\tau\partial_t a= \sigma a - g a^3 -w \partial_x a + \xi^2 \partial^2_x a 
-\frac{1}{\Lambda}\int^x_0 a(x') dx' + \frac{1}{2\Lambda}\int^L_0 a(x')
dx',
\label{eqafring}
\ee      
that must be solved with the boundary conditions $a(L)=-a(0)$, $\partial_x
a(L)=-\partial_x a(0)$. 

Now, we can consider again the linear stability problem, within our
amplitude equation framework. To do that, we write 
$a(x)\sim e^{ikx+\Omega t}$, with $k$ given by (\ref{eq.cabk}), and 
$\Omega$
complex ($\Omega=\Omega_r+i \Omega_i$). Separating into real and
imaginary parts
\bea
&&\tau \Omega_r = \sigma -\xi^2 k^2=\sigma-\frac{\xi^2}{L^2} (2n+1)^2 
\pi^2, 
\label{sigring} \\
&&\tau \Omega_i = \frac{1}{\Lambda k} 
-wk=\frac{L}{\Lambda}\frac{1}{(2n+1)\pi}
-\frac{w}{L}(2n+1)\pi. \label{omegring}
\eea
The onset of instability occurs when $\Omega_r=0$, which results in
$\sigma=\xi^2 k^2$, equivalent to condition (\ref{eq.f'}) when $w/\Lambda\ll
1$.  
Again, we see that intercellular coupling lifts
the degeneracy in the onset of the different modes, since the growth
rate depends on the wavenumber of the mode. Clearly, the fastest
growing mode is that with the largest wavelength, which corresponds to
the mode with a single node. Intercellular coupling also affects the
frequency of the quasiperiodic oscillations. 

To compare Eqs. (\ref{sigring}), (\ref{omegring}) with the
results from the linear stability of the maps in the previous section, 
we can factor out the
rapid oscillations, so $\alpha^n=(-1)^n \beta^n=(-\beta)^n$, and
$\beta^n=e^{\Omega t}=e^{\tau \Omega n}$, by the definition of the
slow time $t=n\tau$. Then, at onset of the instability,
$\Omega=i\Omega_i$, and
\begin{equation}
\alpha=-\beta=-e^{i\tau \Omega_i}\simeq -1 +iwk - \frac{i}{\Lambda k}, 
\end{equation}
which is exactly the expression we had obtained before 
(cf. Eq. (\ref{alfring})). 

\begin{figure}[p]
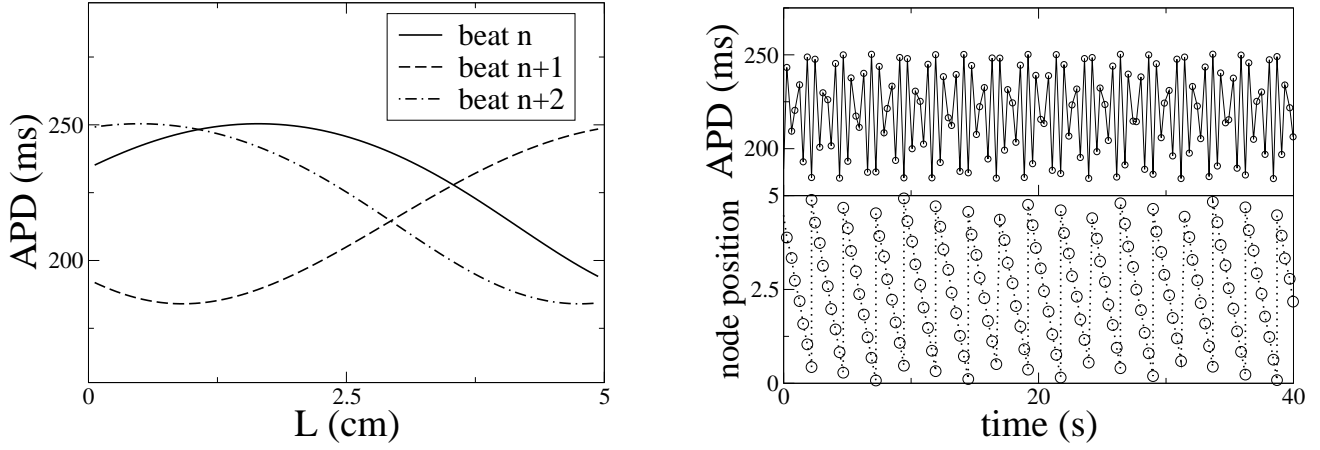

\centerline{
\includegraphics[width=8cm]{fig3a.eps}
\hspace{1cm}
\includegraphics[width=8cm]{fig3b.eps}
}
\vspace{0.5cm}
\caption{Left panel: distribution of APD at consecutive beats, obtained
  solving Eq. (\ref{cable}) in a ring of length $L=5$ cm, for the two-variable
  ionic model. Right panel: evolution in time of the value of the APD at
  $x=L/2$ (top), and position of the node (bottom), defined as the point $x_0$ at 
which
  ${\rm APD}(x_0)=({\rm APD}_{max}+{\rm APD}_{min})/2$.}
\label{fig.ring}
\end{figure}

\begin{figure}[p]
\centerline{
\includegraphics[width=10cm]{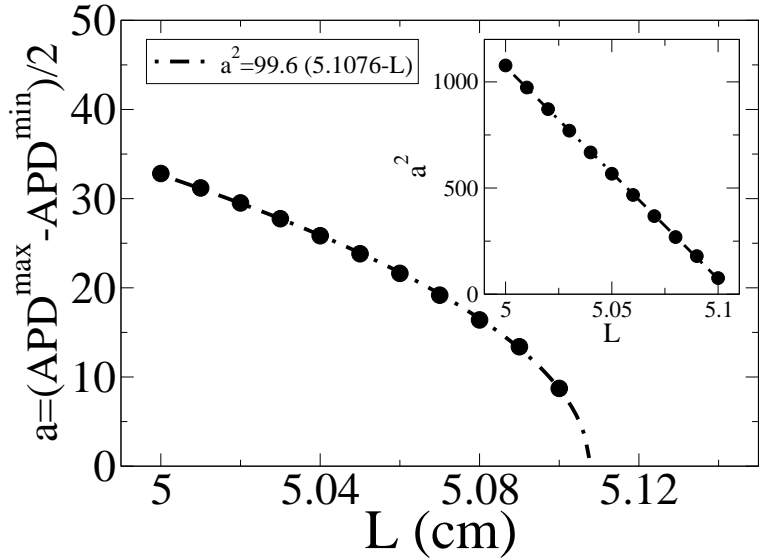}
}
\vspace{0.5cm}
\caption{Bifurcation diagram obtained from simulations of the two-variable
  model in the ring. The bifurcation is supercritical, with a critical ring
  length of $L_c \simeq 5.11$ cm.}
\label{fig.bifring}
\end{figure}

We have checked these results with simulations of the two-variable model. As
expected, the first mode to bifurcate presents only one node, and the
oscillations are quasiperiodic (see Fig. \ref{fig.ring}). The
onset of alternans appears in the simulations for $L_c=5.11$ cm, and the phase
speed of the bifurcating mode is $v=-\Omega_i/k=-1.81\cdot 10^{-3}$ cm/ms, which 
can be
computed from the position of the node as a function of time (see Fig. 
\ref{fig.ring}). 
Using the restitution curves, one obtains that $f'=1$ when $\tau_c=321.5$ ms, for
a conduction velocity of $c=0.0161$ cm/ms. This yields the
prediction $L_c=5.17$ cm. This prediction, however, neglects the
stabilizing effect of the diffusive coupling. 
Setting $k=\pi/L$, and using (\ref{sigring}) with the
parameters of the two-variable model, the 
predicted critical ring length that includes
this effect is $L_c=5.13$ cm, which is in good
agreement with the value $L_c=5.11$ cm in the
numerical simulations (Fig. \ref{fig.bifring}). Then, setting $L=L_c$, the critical
wavenumber is
$k_c=\pi/L_c=0.608 \;\rm{cm}^{-1}$,
 and Eq. (\ref{omegring}) gives $\Omega_i=1.38\cdot
10^{-3}\; \rm{ms}^{-1}$ at the onset of instability, which results in a velocity 
of
the nodes (equal to the
phase velocity) $v=-\Omega_i/k=-2.27\cdot 10^{-3}$ cm/ms. The discrepancy with
the simulation value is due to
$1/(\Lambda k_c)$ corrections. Starting from the full equation
(\ref{beqa}), which does not assume $\Lambda k_c \gg 1$,
it is possible to obtain the modified
expression for the frequency
\begin{equation}
\tau \Omega_i = \frac{\Lambda k}{1+(\Lambda k)^2} -wk.
\end{equation}
This expression gives $v=-1.85\cdot 10^{-3}$ cm/ms, 
in almost perfect agreement with the numerical results.   

Close to onset, the bifurcating solution is therefore generally 
in the form of a traveling wave  
\be
a(x)=\frac{1}{2}(B e^{i(kx+\Omega_i t)}+c.c.),\;\;\; k=\pi/L. 
\ee
Substitution this form in the full nonlinear 
amplitude equation for the ring (\ref{eqafring})
and balancing separately real and imaginary parts, we obtain
at once that the bifurcation is supercritical with a traveling wave
amplitude
\be
|B|=\sqrt{\frac{4}{3}\frac{(\sigma-\pi^2\xi^2/L^2)}{g}}.
\ee 
Given the expected universal validity of the amplitude equation (in the limit
of small dispersion $\Lambda \rightarrow \infty$), this result 
implies that the bifurcation to alternans in the ring will always 
be supercritical if the quasi zero-dimensional bifurcation to alternans in a
periodically stimulated small tissue patch is supercritical, which occurs when 
$g>0$ as in the two-variable model. 
(Note that it is important to distinguish a small tissue patch from an isolated 
cell that has a different restitution curve due to the absence
of diffusive coupling.) We have checked numerically that the bifurcation is supercritical in the
two-variable model (Fig. \ref{fig.bifring}). The amplitude predicted by Eq. (50),
however, is about 60\% higher than the numerically computed amplitude shown in Fig. \ref{fig.bifring}.
This discrepancy can be attributed to $1/(\Lambda k_c)$ corrections that are not small in the
two-variable model due to steep CV-restitution. It should be possible to improve the
agreement by including these corrections in the weakly nonlinear
analysis. Conversely, if $g<0$, the bifurcation in the ring is  
subcritical and the saturation amplitude is generally determined by 
higher order nonlinear terms in the amplitude equation, which can be computed 
analytically if the bifurcation is only weakly subcritical as in the case of the 
Noble model \cite{coeffs}.

\section{Paced tissue\label{paced}}

\begin{figure}[p]
\centerline{
\includegraphics[width=14cm]{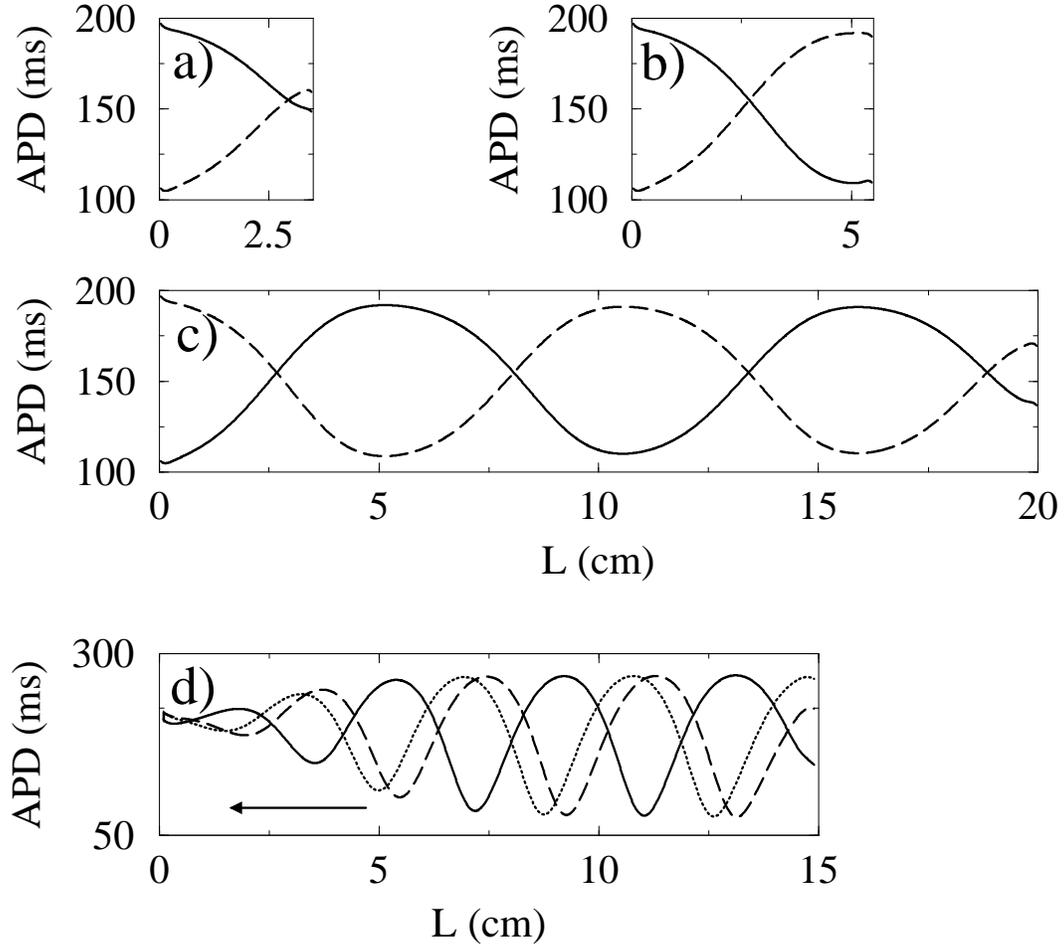}
}
\vspace{0.5cm}
\caption{Distribution of the action potential duration in a long strand 
of tissue obtained simulating Eq. (\ref{cable}) for 
the Noble a), b), and c) and two-variable models, d),
with $\tau=258$ms, and 
$\tau=290$ms, respectively, and several tissue lengths.
The solid and dashed lines correspond to two consecutive  
beats, while the dotted line in d) represents the APD ten beats later.
Thus, in d) the pattern is traveling towards the pacing point and the amplitude 
of alternans grows away from the pacing site as predicted theoretically.}
\label{fig.sim}
\end{figure}

The main difference between the ring and the paced case comes from the 
role
of the boundary conditions. While in the ring the periodic boundary 
conditions result in a quantization condition for the unstable modes, such 
a
condition is absent in the paced case. Thus, the selected wavelength in 
this 
latter case must be related to some intrinsic length scale of the system. 
We show in Fig. \ref{fig.sim} simulations of both the Noble and 
two-variable models. We
have done simulations in long cables to highlight the striking 
spatial regularity of out-of-phase domains of alternans. These patterns suggest
that the formation of discordant alternans is caused by a finite-wavenumber 
linear instability of the basic underlying state, similar to those encountered in
other physico-chemical 
systems, such as Rayleigh-B\'enard convection, Taylor-Couette flow, etc 
\cite{CroHoh}.  
The APD oscillations obtained with the Noble model resemble a 
standing wave (Fig. \ref{fig.sim}c), and those for the two-variable model 
a traveling one (Fig. \ref{fig.sim}d). Thus, at a given point in the
tissue, the 
oscillations in APD are periodic in the first case, and quasiperiodic in
the second. We will see how, within the formalism of the amplitude 
equations, these length scales appear naturally.  

To derive the amplitude equations for paced tissue, we must write the
equivalent of Eqs. (\ref{disp})-(\ref{eq.kernel}) for this case. Now, the
period at a given point will be given by:
\begin{equation}
T^n(x)=\tau + \int_0^x \frac{dx'}{c[{\rm DI}^n(x')]}-\int_0^x
\frac{dx'}{c[{\rm DI}^{n-1}(x')]},
\label{eq.disppaced}
\end{equation}
where $\tau \equiv T^n(0)$ is the period of stimulation at the pacing
point. This simply means that the period of stimulation at a given point in
the tissue is equal to the period of stimulation at the pacing point plus the
difference in arrival time between two consecutive pulses. Notice, however,
that in differential form this equation is the same as in the
ring (cf. Eq. (\ref{eq.dispdiff})). To complete the system, we can use again
Eq. (\ref{eq.kernel}), 
but with a note of care. In fact, in the derivation in the
ring, we use translational symmetry to expand the kernel. In the paced
case this symmetry is broken and one should, in principle, calculate
the kernel for the finite system. When the decay rate of the kernel is
fast this does not seem to be necessary. Eqs. 
(\ref{eq.disppaced})-(\ref{eq.kernel}),
supplemented with non-flux boundary conditions for the oscillations of
APD, give a remarkably good agreement with simulations of the ionic
models, for all the lengths of tissue considered in this paper. It is 
interesting to
notice that the APD itself does not satisfy these boundary conditions, 
as rapid spatial variations in APD (``blips'') 
appear at the two ends of the cable due to the non-flux boundary
conditions for $V$. The amplitude $a$ of alternans obtained by taking
the difference of APD between two consecutive beats,
however, does satisfy very well the non-flux boundary
conditions. This can be checked from
numerical simulations of the ionic models (see Ref. \cite{EcKa02b}). 

It follows that the amplitude equations in the bulk remain the same
[cf. Eq. (\ref{aeqb})], and we only 
have to modify the boundary conditions. 
The condition that the period must
be equal to the pacing interval at $x=0$, $T^n(0)=\tau$, translates
into $b(0)=0$. Then, Eq. (\ref{beqa}) can be solved for $b(x)$ to give
\be
b(x)=\frac{1}{\Lambda}\int_0^x e^{(x'-x)/\Lambda} a(x') dx'.
\label{beqexp}
\ee
As in the previous section, we will assume that the CV-restitution
curve is shallow at the  
bifurcation point, so $\Lambda$ is much larger than the wavelength of the 
pattern ($\Lambda \gg 2\pi/k$). If this is the case then the exponential in
(\ref{beqexp}) can be neglected and the former relation becomes
\be
b(x)\simeq \frac{1}{\Lambda}\int_0^x a(x') dx'.
\label{beq}
\ee
Introducing this into Eq. (\ref{aeqb}), we arrive at the final expression
\be 
 \tau \partial_t a = \sigma a -g a^3  - \frac{1}{\Lambda}\int_0^x a dx'
- w \, \partial_x a+\xi^2 \partial^2_x a.\label{aeqf}
\ee

\begin{figure}[p]
\centerline{
\includegraphics[width=8cm]{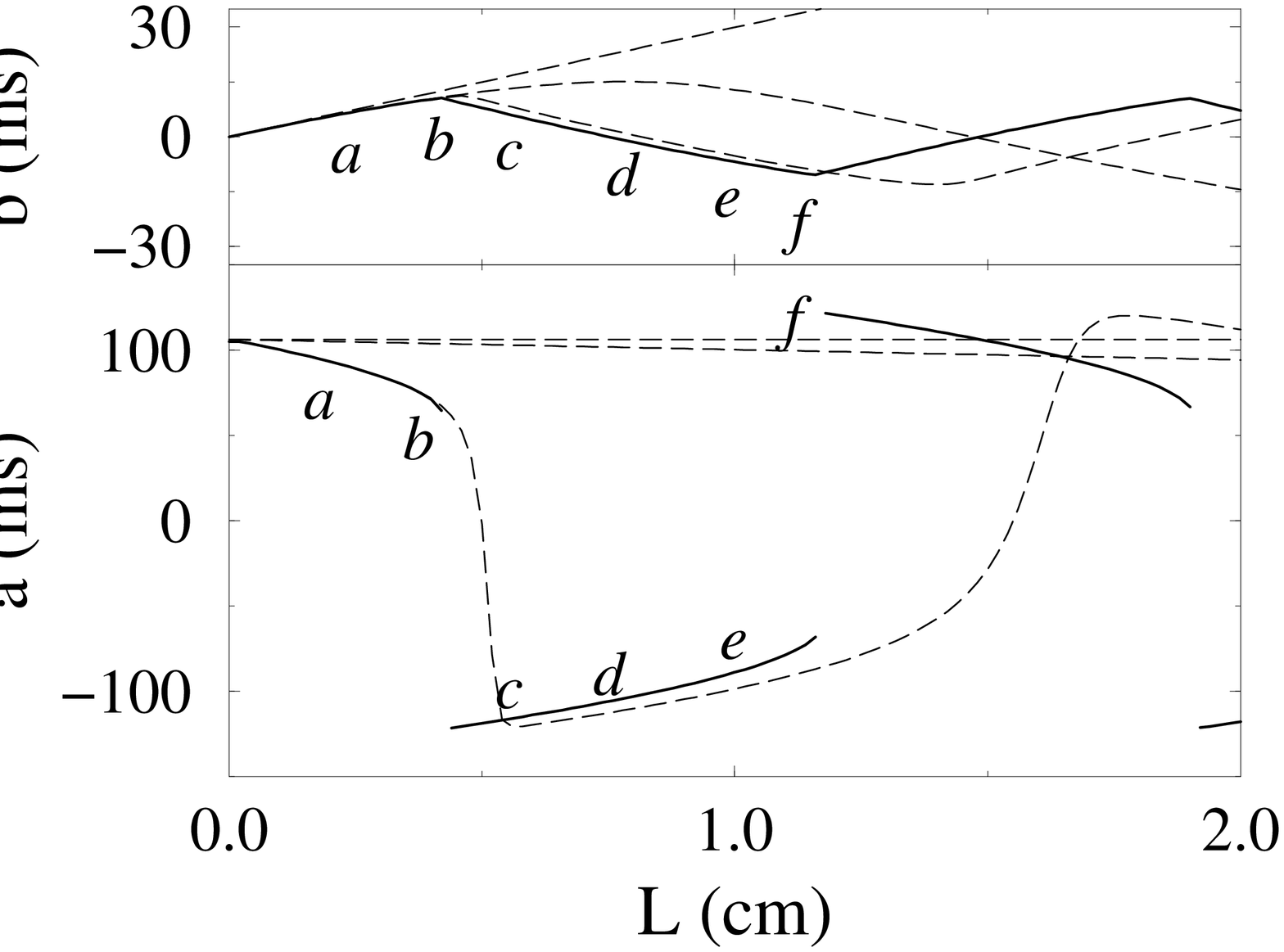}
}
\vspace{0.5cm}
\centerline{
\includegraphics[width=10cm]{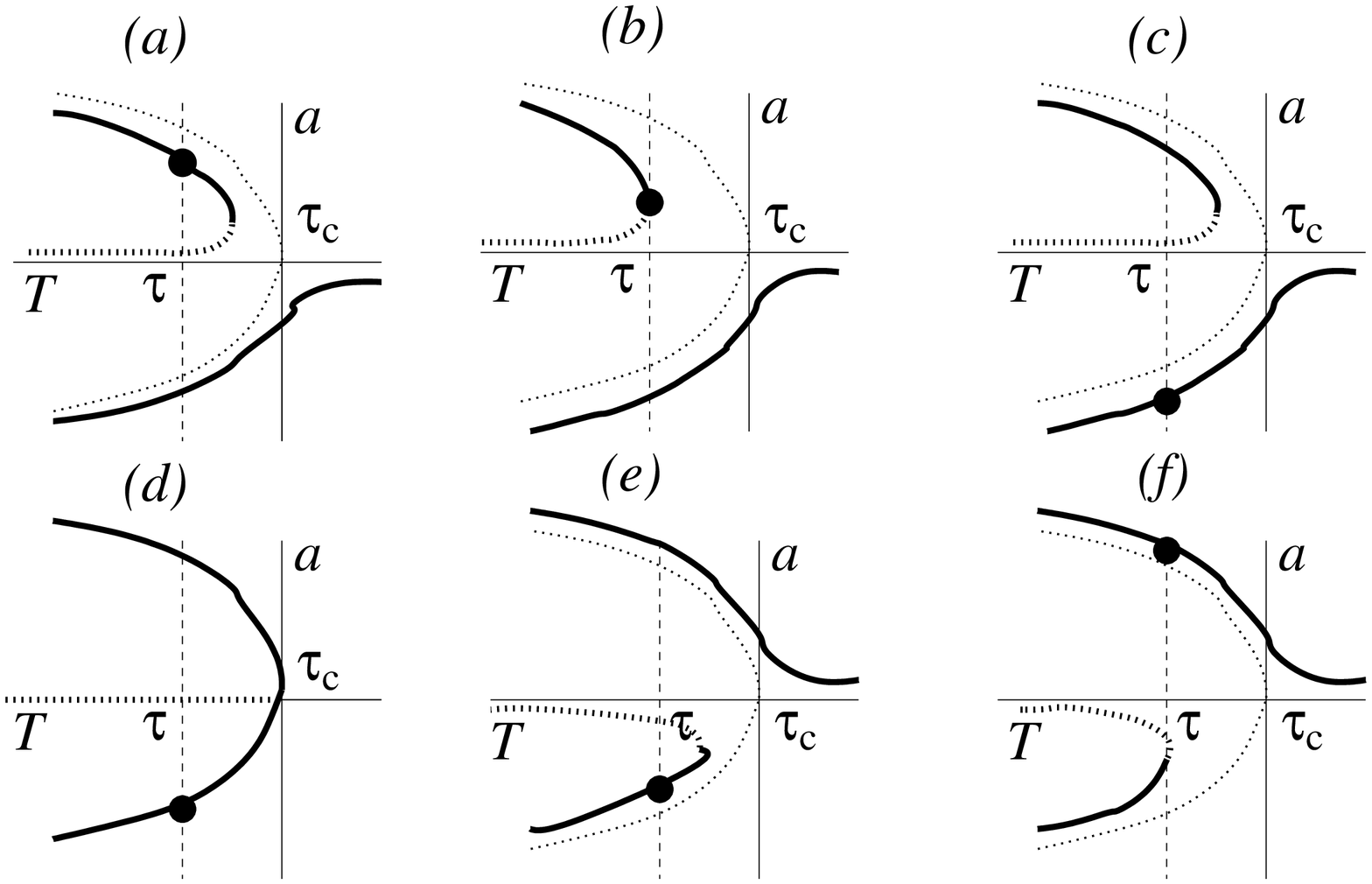}
}

\caption{Upper panel: Oscillations in action potential 
duration and period, obtained
simulating Eqs. (\ref{beq}) and (\ref{aeq}), with the parameters of the 
two-variable model, 
and $\tau=290$ms. Starting with a constant value of $a$, we show in dashed
lines the evolution of the system at times $t=0,10\tau,50\tau$, and
$500\tau$.
The final state is shown in solid lines.
Below we sketch the bifurcation diagram at several points along the 
cable for this final state. The dashed line denotes
the value of $T(0)=\tau$ and the filled circle the value of the amplitude 
of 
oscillation in APD in each case. As we go along the tissue
$b$ becomes positive and the pitchfork (dotted line) 
becomes and imperfect bifurcation
[cf. Eq. (\ref{aeq})] (a). At point (b) the saddle-node is at $T=T(0)$.
There is a jump to the other phase and $b$ starts to decrease (c). At
(d) $b=0$ and a perfect pitchfork bifurcation is recovered, after which 
the branches switch place (e). Finally, the saddle-node reaches $T=T(0)$ 
and
the process starts again.}
\label{fig.bif}
\end{figure}

With the help of equation (\ref{aeqf}) the formation of the nodes 
is easy to explain. Let us first neglect
the influence of the electrical coupling between cells on the
APD (that later on will be shown to be crucial), and consider the
equation
\be
 \tau \partial_t a = \sigma a -g a^3  - b. \label{aeq}
\ee 
Starting at $x=0$ with a 
constant amplitude of oscillation for the APD $a_0$, 
the oscillation in the period $b$ 
will increase along the tissue [cf. Eq. (\ref{beq})], 
effectively decreasing the 
value of $a$, until it crosses zero and goes into the branch with opposite 
phase (see Fig. \ref{fig.bif}). Thus, the system 
tends to create discordant alternans spontaneously, through 
CV-restitution.
However, without the derivative terms the gradients become ever steeper, 
tending to a discontinuous limit as $t \rightarrow \infty$.
To see this, we can differentiate Eq. (\ref{aeq}), to obtain the steady 
state 
equation for the slope of the oscillations in APD 
$\partial_x a=a/\Lambda(\sigma-3ga^2)$, that can be integrated to obtain
$x=\Lambda [\sigma \ln (a/a_0)-3 g (a^2-a^2_0)/2]$, with
$a_0=a(0)=\sqrt{\sigma/g}$. 
When $a^2(x_0)=\sigma/3g$ the slope becomes infinite, denoting the
formation of a singularity at $x_0$. 
The singular behavior originates from the use of the local
APD-restitution relation [Eq. (\ref{res})] that allows two
nearby cells to have different APDs, whereas this is prevented
in the cable equation by the spatial diffusive coupling.
The distance between singularities can be obtained
integrating Eq. (\ref{aeq}) from $a^2=4\sigma/3g$ (the value after the singularity) to
$a^2=\sigma/3g$ (the value at the next singularity). Then, we obtain the
wavelength (as twice the distance between singularities): $\lambda=2\Lambda
\sigma [3/2 - \ln(2)]$, that does not correspond to the  
right length scale for this problem (compare Figs. \ref{fig.bif} 
and \ref{fig.sim}). Once the derivative terms are included, 
the agreement between the ionic 
models and the amplitude-equations becomes very good.
To check this point, we have derived the coefficients $w$ and $\xi^2$
from the restitution curves of the Noble model (in the two-variable
model their values are given by Eqs. (\ref{w2v}) and (\ref{xi2v})).

\subsection{Measurement of the coefficients $w$ and $\xi$ in the amplitude
  equation.} 

\begin{figure}[p]

\centerline{
\includegraphics[width=12cm]{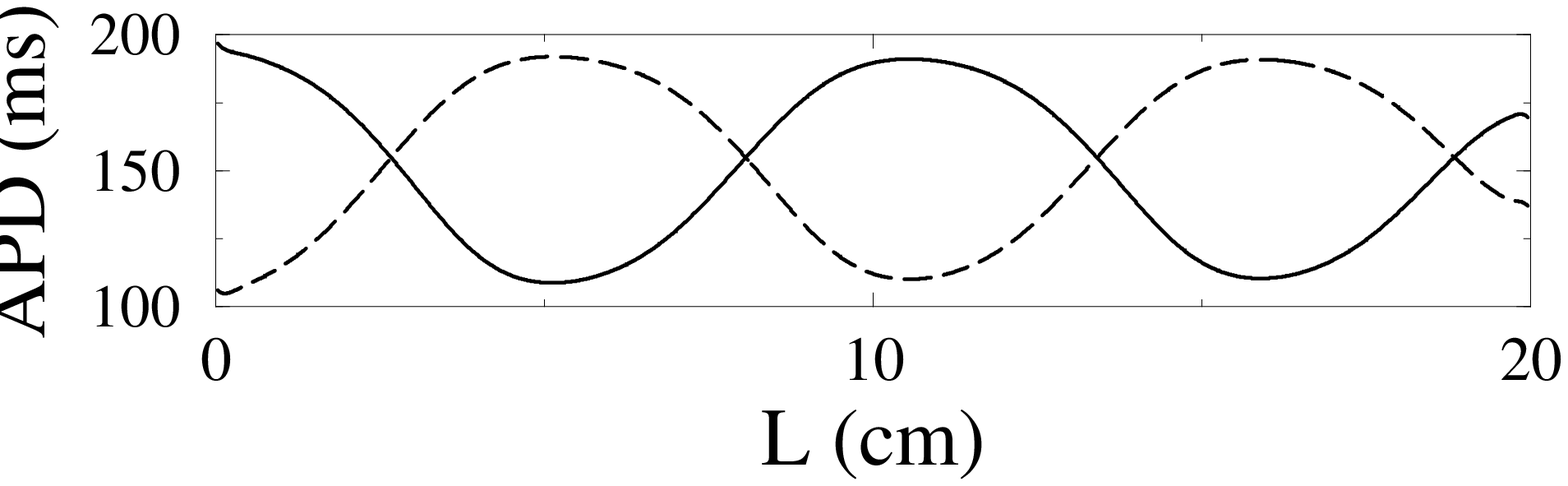}
}
\centerline{
\includegraphics[width=12cm]{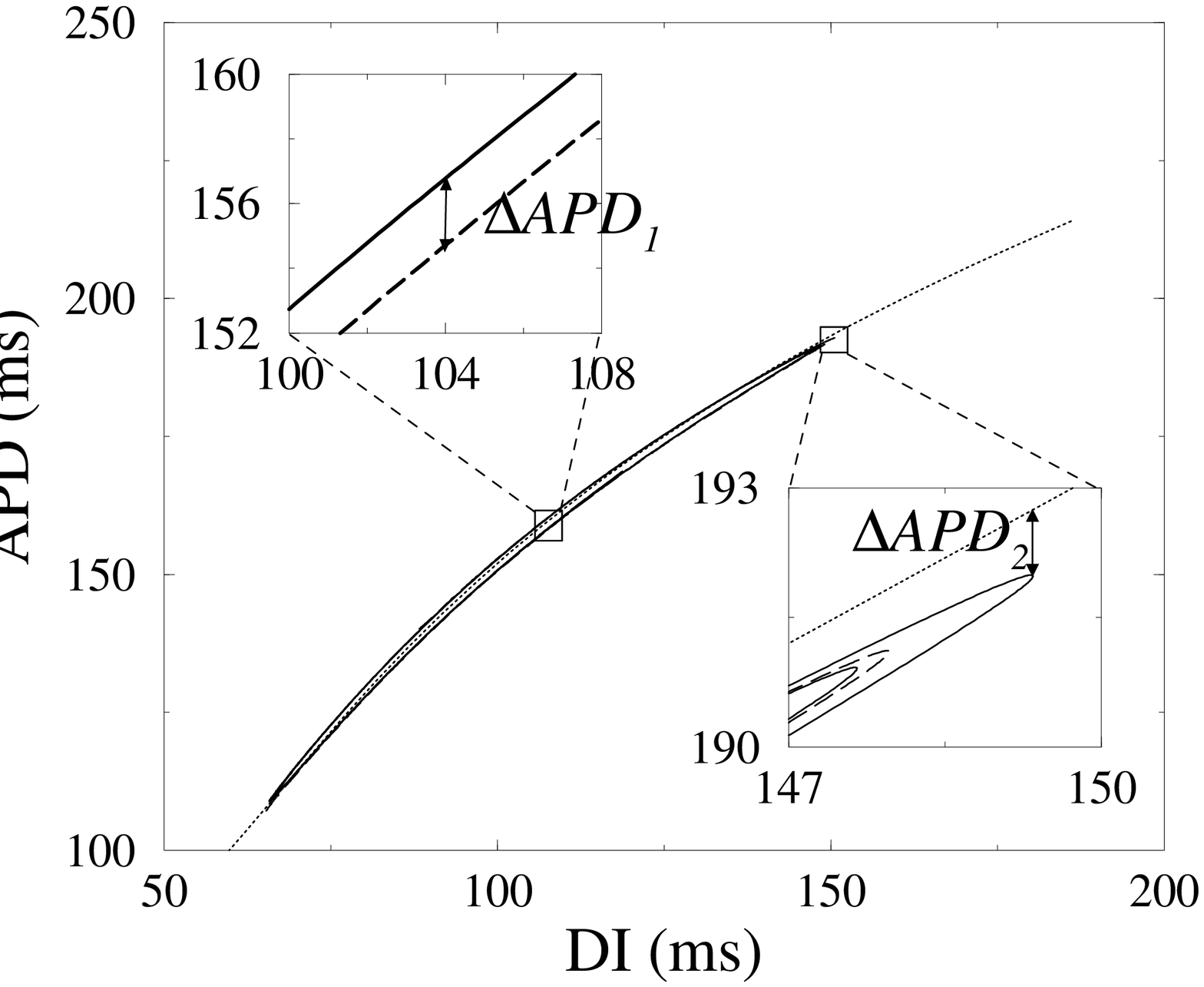}
}
\vspace{0.5cm}
\caption{
a) Distribution of the action potential potential at two consecutive 
beats in a long strand of tissue obtained simulating Eq. (\ref{cable}) 
for 
the Noble model, with $\tau=258$ms. 
b) 
Restitution curves for the Noble model. The dotted line corresponds to that 
obtained using a S1-S2 protocol, while the solid and dashed lines are the 
restitution curves corresponding to the two beats in a). The difference 
between these two latter curves at a point $x^*$ corresponding to a node in
DI, $\Delta {\rm APD}_1={\rm APD}^{n+1}(x^*)-{\rm APD}^{n}(x^*)$, gives the change in APD 
corresponding to a negative or positive slope. The value $\Delta
{\rm APD}_2$ is obtained as the difference between the maximum value of the
APD in a), and the value of APD 
for the same DI obtained using a S1-S2 protocol, where no gradients
are present.  
\label{fig.calcoeffs}}
\end{figure}

\begin{figure}[p]
\centerline{
\includegraphics[width=10cm]{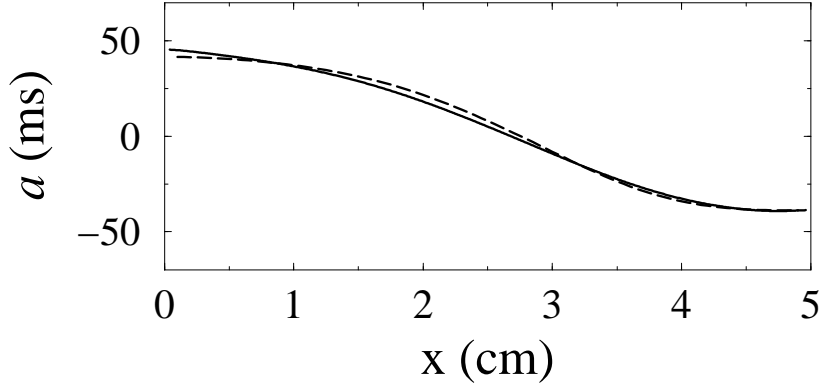}
}
\vspace{0.5cm}
\caption{Comparison of the results obtained simulating the Noble model 
(solid line) 
and the amplitude equation (\ref{aeqf}) (dashed line), with the 
coefficients 
given in Table \ref{lengths}.
\label{fig.Nobcomp}}
\end{figure}

To measure the coefficients $w$ and $\xi$ we will consider a case where tissue
is paced at a constant period, for values close to, but beyond, the onset of
alternans. If we plot the map ${\rm APD}^{n+1}$ in terms of ${\rm DI}^{n}$ corresponding
to those simulations (Fig. \ref{fig.calcoeffs}) we obtain a restitution curve 
that
differs from the one obtained using a S1-S2 protocol, where DI is spatially
uniform. During discordant 
alternans the electrotonic currents generated by
the spatial modulation of DI along the cable modify the cable restitution curve 
computed in Appendix
\ref{ap2var} for a spatially uniform DI, which is equivalent to
the one obtained using a S1-S2 protocol, $f_{S1S2}$ (see
Fig. \ref{fig.rest}). The cable restitution curve with a 
spatially modulated DI, which we define as $f_{mod}$, can be different on even
and odd beats during discordant alternans because the sign of the
spatial gradient of DI in the nodal region alternates from beat to beat.
If the system presents no
memory, then the difference between $f_{mod}$ and $f_{S1S2}$ is due to the
presence of gradients of DI. Expanding the kernel 
in Eq. (\ref{eq.kernel}), we can write: 
\begin{equation}
{\rm APD}^{n+1}(x)=f_{S1S2} [{\rm DI}^n(x)]-w \partial_x {\rm DI}^n(x) +\xi^2 \partial^2_x 
{\rm DI}^n(x)\equiv
f_{mod}[{\rm DI}^{n}(x)]. 
\end{equation}
Let us now consider a point in space $x^{*}$ where there is a node in DI, so 
${\rm DI}^{n+1}(x^*)={\rm DI}^{n}(x^*)={\rm DI}^{*}$. Then, the S1-S2 restitution curve  
gives the same value at two consecutive beats, so the split in the restitution 
curve in Fig. \ref{fig.calcoeffs} will be due to the gradients. Given that the
alternans profile in Fig. \ref{fig.calcoeffs} is, to a good approximation,
sinusoidal, then at that point
it is also satisfied that $\partial^2_x  
{\rm DI}^n(x^*)\simeq \partial^2_x {\rm DI}^{n+1}(x^*)\simeq 0$. Subtracting the APD at consecutive beats 
we have:
\begin{equation}
\Delta {\rm APD}_1 \equiv f_{mod}[{\rm DI}^{n+1}(x^*)] - f_{mod}[{\rm DI}^{n}(x^*)]= w [ \partial_x 
{\rm DI}^{n}(x^*) -  
\partial_x {\rm DI}^{n+1}(x^*) ].
\end{equation}
Then, we have 
\be
w\simeq\frac{\Delta {\rm APD}_1}{\partial_x {\rm DI}^{n}(x^*)-\partial_x {\rm DI}^{n+1}(x^*)}.
\ee
To calculate the coefficient $\xi$, we just consider the point of tissue
$x^{max}$ where $f_{mod}$ has a local maximum, corresponding to an
antinode of alternans. At that point
$\partial_x {\rm DI}^n(x^{max})=0$, and then: 
\begin{equation}
f_{mod}[{\rm DI}^n(x^{max})]=f_{S1S2}[{\rm DI}^n(x^{max})] + \xi^2
\partial^2_x {\rm DI}^n 
(x^{max}).
\end{equation}
In this case $\xi$ can be obtained from the difference between the S1S2
restitution curve and the one obtained during discordant alternans. Defining 
$\Delta {\rm APD}_2 \equiv 
f_{mod}[{\rm DI}^n(x^{max})] - f_{S1S2}[{\rm DI}^n(x^{max})]$, we have:
\begin{equation}
\xi^2\simeq\frac{\Delta {\rm APD}_2}{\partial^2_x {\rm DI}^n(x^{max})}.
\end{equation}
In Table \ref{lengths} we show the values of the coefficients calculated 
in 
this manner. The comparison between simulations of the amplitude 
equation (\ref{aeqf})
using these coefficients and the Noble equations is very good, as shown
in Fig. \ref{fig.Nobcomp}. 

\begin{table}[t]
\begin{tabular}{|l|c|c|c|c|c|c|}
\hline
 Model & $\Lambda$ & $w$ & $\xi$ & $\lambda_{theor}/4$ & $\lambda_{sim}/4$ & 
$L_{min}$ 
 \\  \hline 
Noble &  49.1  & 0.045  & 0.18  & 2.33 & 2.6 & 2.75 \\
Two-variable &  3.55  &  0.031  &  0.235 & 1.33 & 1.1 &  1.15 \\ \hline

\end{tabular}
\caption{Values of the parameters in the amplitude equation (\ref{eq.aintro})
  associated with different length scales, and comparison of the simulated and
  theoretical wavelengths for the unstable modes in the paced case (all distances
  in cm). The
  coefficients $w$ and $\xi$ are obtained analytically from
  Eqs. (\ref{w2v})-(\ref{xi2v}) in the two-variable model, and numerically
  from the APD-restitution curves as 
  explained in the text, for Noble model. The values of $\Lambda$ are obtained
from the CV-restitution curves computed numerically, in both cases. The 
wavelength
  $\lambda_{theor}$ is predicted from Eq. \protect(\ref{lsta}), for Noble, 
or Eq. \protect(\ref{ltra}), for the two-variable model (see text for
  details), while $\lambda_{sim}$ is
  obtained from simulations
of Eq. \protect(\ref{cable}) in a long cable. These wavelengths give also a
  good estimate of the minimal length of tissue at which a node
  appears.\label{lengths}} 
\end{table}

\subsection{Linear stability analysis}

Although nonlinear effects determine the saturated amplitude
of alternans, the spacing between nodes of discordant alternans, 
and the velocity of the nodes in the case of traveling waves, are well
predicted by linear stability theory close to the bifurcation.  

The genesis of discordant alternans can be understood  
by computing the linear stability spectrum
of the spatially homogeneous
state ($a=0$). We have calculated this 
spectrum numerically for different values of $L$, 
and compared it with the analytical results obtained 
in the large $L$ limit. The main result 
is that the wave pattern can emerge from the amplification of
\emph{either} a unique finite wavelength 
mode, which yields a stationary pattern,  
\emph{or} from a discrete set of  
complex modes that approach a continuum in the 
limit of large $L$, and yields a traveling pattern 
(see Fig. \ref{fig.sim}). 
There is indeed experimental evidence for both stationary
\cite{Pasetal} and traveling \cite{Foxetal} waves.

\begin{figure}[p]
\centerline{
\includegraphics[width=14cm]{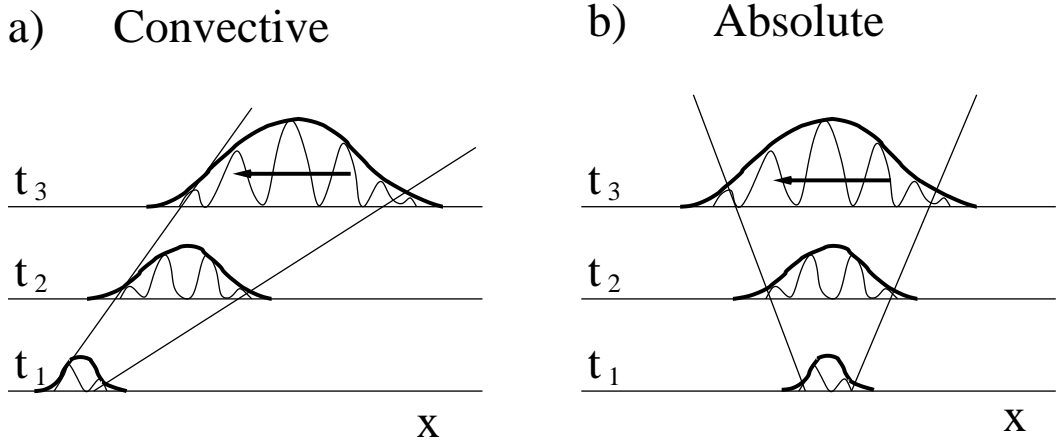}
}
\caption{Convective {\em vs} absolute instability. When the instability is
convective (as in a)), an initial localized perturbation is advected as
it grows (here $t_3>t_2>t_1$), so at a given point in space it decays. 
The instability becomes absolute when the wavepacket grows at any given 
point 
in space (b). This is signaled by a vanishing group velocity. It should be 
noted
that the phase velocity, however, does not vanish and, in general, it does
not have to be in the same direction as the group velocity.}
\label{conv}
\end{figure}

In the large $L$ limit it is possible to obtain analytical expressions for 
the wavenumber and onset of the waves, using the dispersion related associated
to Eq. (\ref{aeqf}). Differentiating this equation  
with respect to $x$ and letting $a(x,t) \sim e^{ikx+\Omega t/\tau}$,
with both $\Omega\equiv \Omega_r+i\Omega_i$  
and $k\equiv k_r+ik_i$ complex,
yields at once 
\begin{equation}
\Omega=\sigma-\xi^2k^2-i\left[wk-\frac{1}{\Lambda k} \right].\label{spec}
\end{equation}
Except for reflection symmetric systems (i.e. system invariant
under the transformation $x \rightarrow -x$), or
with periodic boundary conditions that impose $k \in {\cal R}e$, the
wavenumber will be complex. An instability occurs if there exists a complex
wavenumber $k$ for which $\Omega_r >0$. It may happen, however, that the
unstable mode has a non-zero group velocity $\partial \Omega/\partial k \neq
0$. Then, the instability will grow in the frame moving with the group
velocity, but decay at a fixed position. This is the signature of a 
convective instability \cite{CroHoh} where perturbations 
are transported as they grow, similarly to, for instance,
Taylor-Couette vortices developing in an axial flow \cite{Babetal}.
In such a situation, patterns are only transient in a finite system (unless
a constant forcing is provided), since they disappear through the boundaries
in finite time. A sustained instability develops if the group velocity of the 
unstable mode is
zero $\partial \Omega/\partial k=0$, in which case the instability grows at a
fixed location in space. This is known as an absolute instability
(Fig. \ref{conv}). The condition
\be
\frac{d\Omega}{dk}=0=-2\xi^2 k -iw -\frac{i}{\Lambda k^2}
\label{domega}
\ee 
yields a prediction for the complex wavenumber corresponding
to the onset of absolute instability.
In the limit $w\rightarrow 0$, it becomes simply 
 \be
 k=\pm \frac{\sqrt{3}}{2(2\xi^2 \Lambda)^{1/3}}-
 \frac{i}{2(2\xi^2 \Lambda)^{1/3}},\label{wavenumberabs}
 \ee
 and corresponds to modes growing exponentially in space. This mode is
 evident in Fig. \ref{fig.sim}d, where the amplitude grows away from the pacing 
 point before saturating due to nonlinear effects. There exists a third
 solution of Eq. (\ref{domega}) that gives a purely imaginary wavenumber (an
 exponentially 
 decaying mode without sinusoidal part). It can be checked, however, that this
 mode belongs to the spectrum associated with the equation resulting from
 differentiating Eq. (\ref{aeqf})   
with respect to $x$, but not to Eq. (\ref{aeqf}) itself. 

 Substituting expression (\ref{wavenumberabs}) into Eq. (\ref{spec}), 
 we deduce that the threshold of absolute instability occurs
 when 
\begin{equation}
\sigma_{th}=(3/2)(\xi/2\Lambda)^{2/3},\label{onset.2var}
\end{equation}
 with a pattern of wavelength
 \begin{equation}
 \lambda=\frac{4\pi}{\sqrt{3}}(2\xi^2 \Lambda)^{1/3}.\label{ltra}
 \end{equation}
This wavelength 
 agrees well with that observed in simulations of the two-variable
 model (see Table \ref{lengths}). The frequency is, in this approximation,
 $\Omega_i=(3\sqrt{3}/2)(\xi/2\Lambda)^{2/3}$, resulting in $\Omega_i\simeq
 0.27$ for the two-variable model, and
 the pattern 
 travels with phase velocity $v_{ph}=-\Omega_i/k_r$. 

We have confirmed the validity of the former analysis by numerically solving
the linear eigenvalue problem associated 
with (\ref{aeqf}). To obtain the linear spectrum for a given value of $L$, we
have linearized 
and discretized in space Eq. (\ref{aeqf}), using a finite difference 
representation of the derivatives and the trapezoidal rule for the 
integral.
Looking for exponentially growing or decaying solutions 
$a_i(t) \sim a_i e^{\Omega t/\tau}$, we obtain a set of $N$ coupled linear 
algebraic equations
\be
\Omega a_i = \sigma a_i -\frac{w}{2dx} (a_{i+1}-a_{i-1}) + 
\frac{\xi^2}{dx^2} (a_{i+1}+a_{i-1}-2a_i)
-\frac{dx}{\Lambda} \sum^{i-1}_{j=0} 
\frac{1}{2}(a_j+a_{j+1}),\;\;\;i=1\dots N,
\label{aeqdisp}
\ee
where $L=N dx$, and we typically use $dx=0.05$ cm. The non-flux boundary 
conditions now become $a_0=a_2$, and $a_{N+1}=a_{N-1}$. The 
resulting eigenvalue problem is then solved for the complex growth rate
$\Omega=\Omega_r+i\Omega_i$, and
the corresponding eigenmodes. These will be either real (with $\Omega_i=0$),
in which case we will talk about stationary modes, or come in complex
conjugate pairs, resulting in traveling waves. In order to compare with the
analytical predictions, we calculate the wavenumber from the number of nodes
$n$ of the eigenmodes (of both their real and imaginary parts), as $k_r\simeq
\pi n/L$. 

\begin{figure}[p]
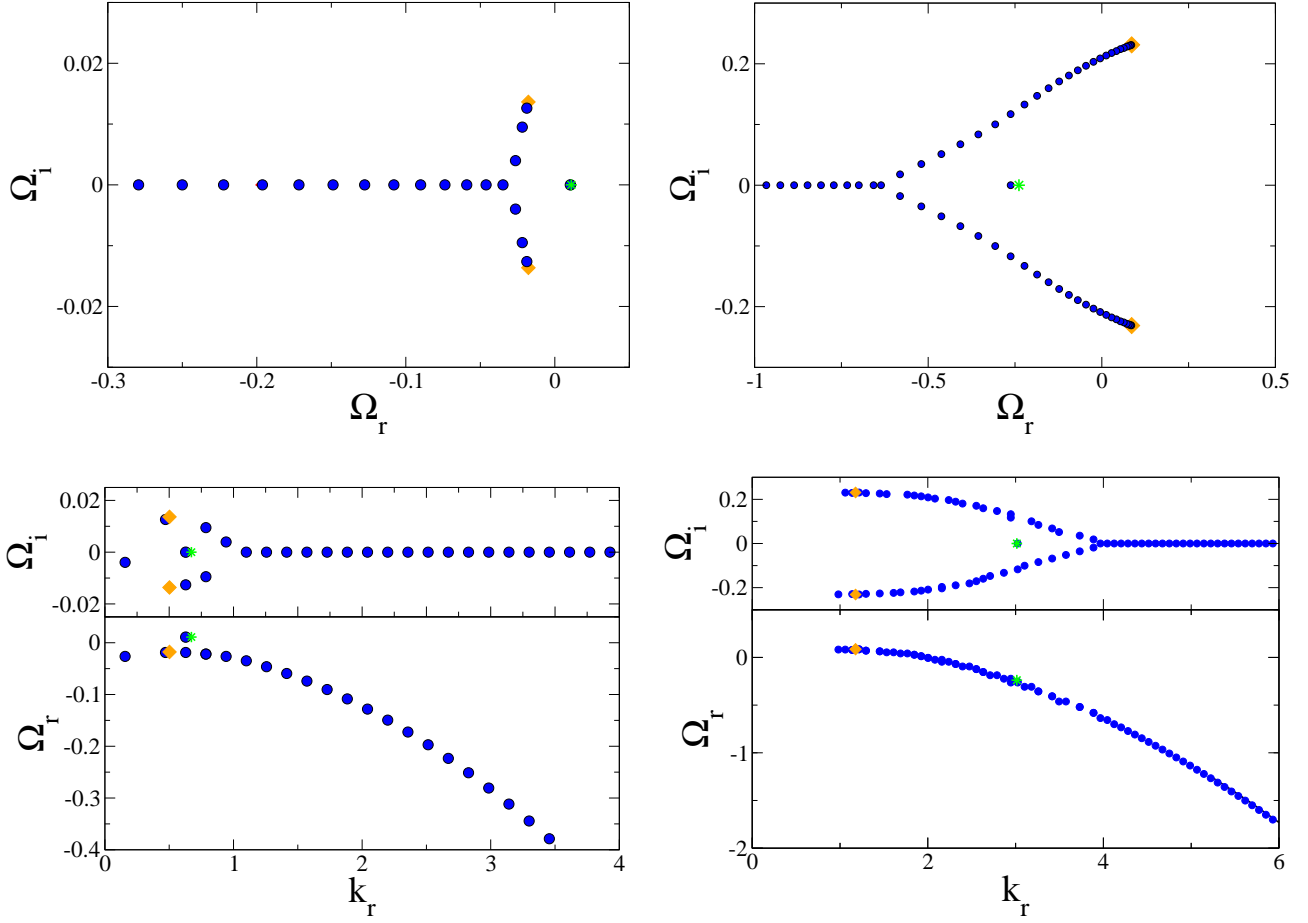

\centerline{
\includegraphics[width=8.25cm,clip=true]{fig10a.eps}
\hspace{0.25cm}
\includegraphics[width=8.25cm,clip=true]{fig10b.eps}}
\vspace{0.5cm}
\centerline{
\includegraphics[width=8.25cm,clip=true]{fig10c.eps}
\hspace{0.25cm}
\includegraphics[width=8.25cm,clip=true]{fig10d.eps}
}
\caption{(Color online). Top panels: comparison between the spectra obtained solving
  the linear eigenvalue problem given by Eq. (\ref{aeqdisp}) (circles) and the
  analytical predictions (diamonds and stars),    
  for the parameters of Noble
  model (left) at a pacing period of $\tau=258$ ms and $L=20$ cm, and the 
two-variable model
  (right), with $\tau=290$ ms and $L=40$ cm. Bottom panels: dependence of growth 
rate and frequency on
  the real wavenumber. For the linear
  eigenvalue problem (circles), the wavenumber has been calculated from
  the number of nodes $n$ of the eigenmodes $k_r \simeq n\pi/L$. The star
  corresponds to the stationary mode, with $k_r=(w\Lambda)^{-1/2}$,
  $\Omega_i=0$, and $\Omega_r=\sigma-\xi^2/(w\Lambda)$. The diamonds
  correspond to the results for the standing waves, obtained solving Eq. 
(\ref{domega}). For the
  parameters of Noble (left) the first 
  mode to bifurcate is an stationary mode, corresponding to the isolated
  eigenvalue. For the two-variable model (right), there is a continuous branch
  of modes that bifurcates first.  }
\label{fig.linstab}
\end{figure}

From Fig. \ref{fig.linstab} it can be seen that the spectrum consists of a 
continuous
branch, plus an isolated eigenvalue, whose origin is
probably due to the boundary conditions. This can be
motivated by noting that 
$\cos k_rx$ is 
an exact eigenvector of Eq. (\ref{aeqf}) linearized around
$a=0$ that satisfies  $\partial_x a=0$ at the two cable ends. Substituting
that solution into Eq. (\ref{aeqf}), we obtain $\Omega_i=0$,
$\Omega_r=\sigma-\xi^2 k^2_r$, and a wavelength
$\lambda=2\pi/k_r$ given by  
\be 
\lambda = 2\pi (w\Lambda)^{1/2}.\label{lsta}
\ee
This is in good agreement with the wavelength of standing waves observed
in simulations of the cable-Noble equation
(Table \ref{lengths}). Although this exact mode only exists for given specific
values of the length of the system (when $L$ is an integer multiple of
$\lambda/2$), the computed spectrum shows that it also persists for arbitrary
values of $L$. The onset of standing waves then is predicted to occur at
\begin{equation}
\sigma_{th}=\xi^2/(w\Lambda).\label{onset.Nob}
\end{equation}

\begin{figure}[p]
\centerline{
\includegraphics[width=12cm]{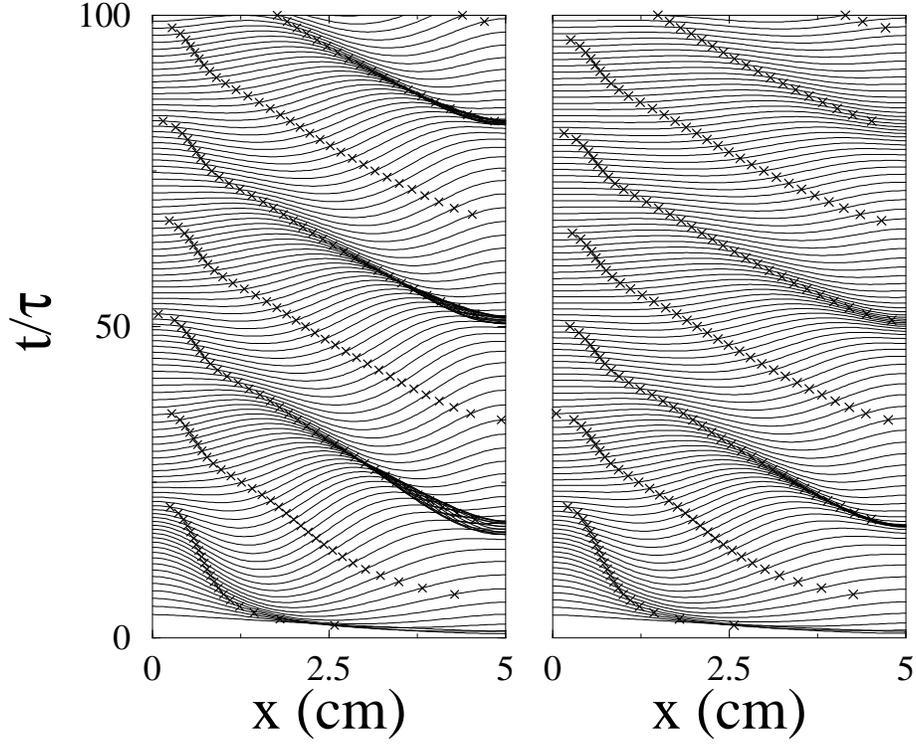}
}
\caption{Space-time plots of $a$  
obtained by simulations of Eq. \protect(\ref{aeqf}) for
parameters of the two-variable model, showing   
absolutely unstable (left and $\tau=295$ ms) and  
convectively unstable (right and $\tau=298$ ms ) wave patterns. The crosses 
denote
the positions of the nodes ($a=0$).
}
\label{examples}
\end{figure}

 The continuous spectrum ends in the branch points given by the absolute
  instability, with both $k$ and $\Omega$
  complex, corresponding to traveling waves. Then, the bifurcating modes can
  be either stationary or 
  traveling waves, depending if it is the isolated eigenvalues, or the branch
  points, that bifurcate first. In the limit $w \rightarrow 0$ the
  growth rate of the standing wave $\Omega_r
  \rightarrow -\infty$, and the resulting pattern will always be that of
  traveling waves. There is, therefore, a critical value of $w$ below which
  the isolated eigenvalue presents a lower 
 growth rate than the branch points, and complex modes that grow exponentially
  at  
 large $x$ are the most unstable. This is the case for the two-variable
  model (see Fig. \ref{fig.linstab}), where the modes pass from being
  convectively to absolutely unstable, as the pacing rate is decreased (Fig. 
\ref{examples}).
From Eqs. (\ref{onset.2var}) and (\ref{onset.Nob}) it can be deduced that
 traveling waves are favored over stationary waves
  when $\Lambda=c^2/(2c') \ll \xi^4/w^3$,
 and hence for strong CV-restitution,
 and vice versa for weak CV-restitution. 

 Thus, our results demonstrate that the formation
 of discordant alternans is crucially affected by the 
 effect of electrical coupling (diffusion) on repolarization,
 in addition to restitution and dispersion. 
 Dispersion is responsible for the formation of nodes and
 spatial gradients of APD that steepen with time. 
 Diffusion, in turn, tends to spread the APD spatially,
 and also induces a drift of the pattern away from the pacing site
 that is induced by the more subtle gradient term 
 ($-w\partial_x a$) in the amplitude equation. When
 dispersion is sufficiently weak, it may be balanced by drift
 and produce a stationary pattern. In the opposite limit,
 the tendency for dispersion to form steep gradients
 of APD is balanced by the spreading effect of diffusion. 
 Nodes then travel, cyclically 
 disappearing (appearing) at the pacing (opposite)
 end of the cable.  

The length scale for the standing
waves in the paced case is the same as that
for the strictly periodic motion in the ring [see Eq. 
(\ref{ring.period})]. 
In the former case, however, the non-flux boundary conditions 
pinned the node (imposed $\Omega_i=0$), 
and that was the only possible wavelength for a bounded state. The
only possibility of having a different wavelength, and a traveling node, 
was to pick up an exponential 
part ($k$ complex), which is precluded in the ring by the boundary 
conditions.
There, the wavelength is selected by the boundary 
conditions, and depends on the length of the system. This, in turn, 
selects the velocity of the node through Eq. (\ref{omegring}), that in
general will be different from zero. 
Comparing the critical length scales in both cases, we see that for
the mode with largest wavelength in the ring ($n=0$), this 
results in a critical tissue length of $L_c=\lambda/2=5.78$ cm, for Noble, 
and
$L_c=\lambda/2=5.17$ cm for the two-variable model. Thus, the length scale 
in the ring is comparable with the one in the paced case for the Noble
model, while for the two-variable model it is significantly larger (see 
Table \ref{lengths}).

\subsection{Stability diagrams}

\begin{figure}[p]
\centerline{
\includegraphics[width=8cm]{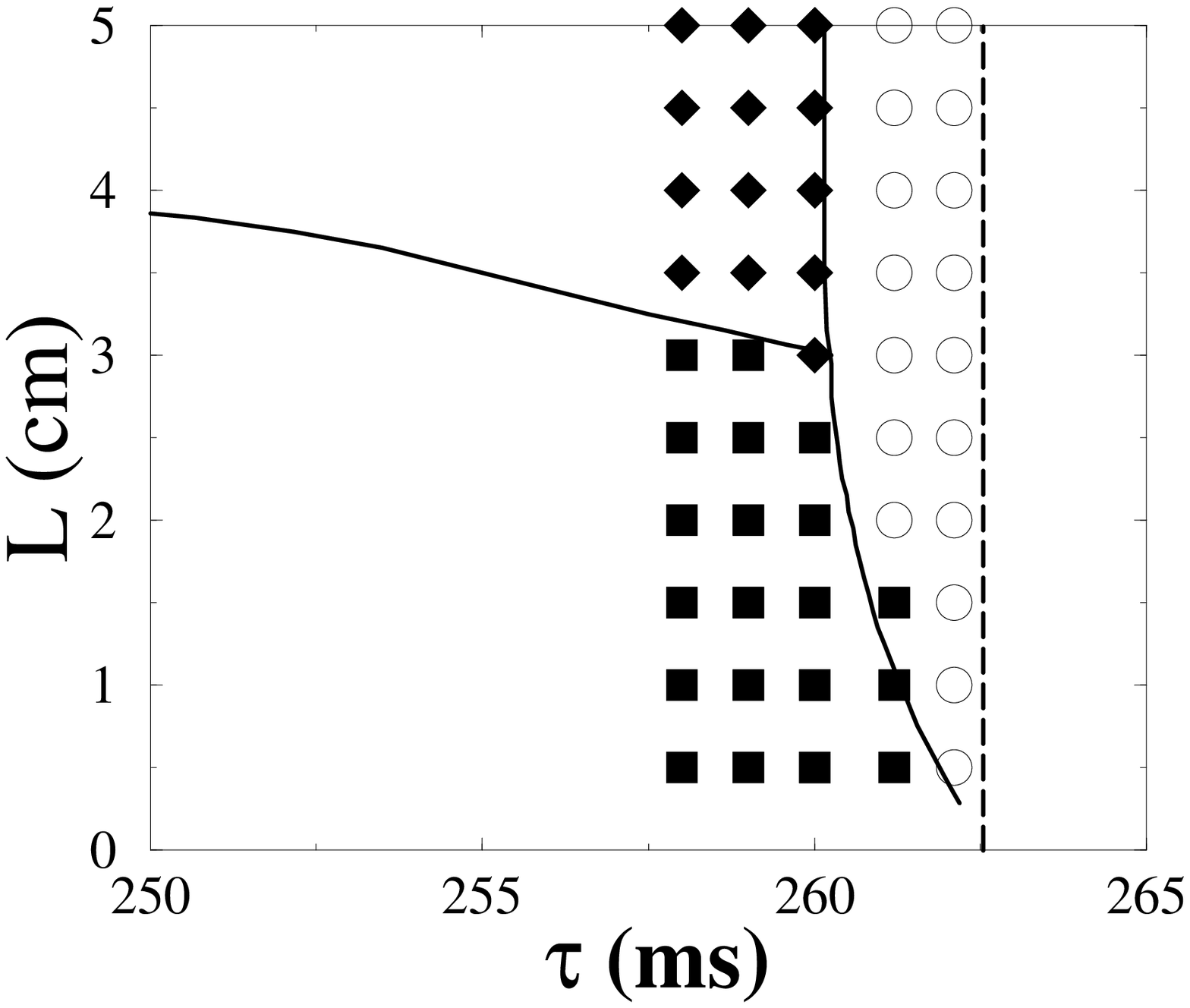}
\includegraphics[width=8cm]{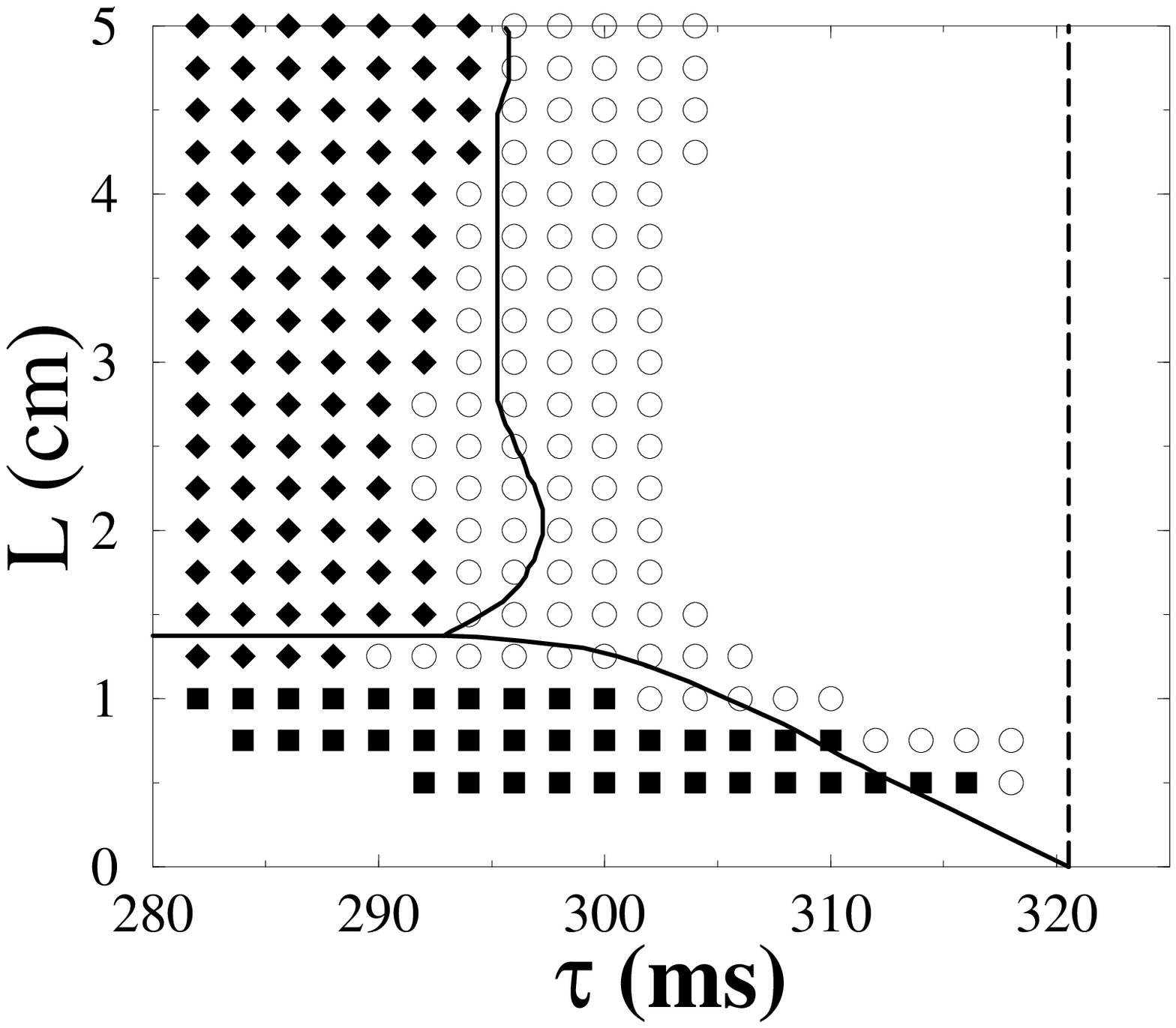}
}
\vspace{0.2 cm}
\caption{Stability diagram of the Noble (left panel) and
two-variable (right panel) cable models with domains of no-alternans
(open circles), concordant alternans
(filled squares), and discordant alternans
(filled diamonds); conduction blocks form
at smaller $\tau$ not shown here. Boundaries between the 
same domains obtained by simulations
of the amplitude equation (\protect \ref{aeqf} )
are shown by solid lines. The dashed line denotes the bifurcation period
for alternans predicted by the map of Eq. (\ref{res}).}
\label{phasediag}
\end{figure}

To compare with previous numerical and experimental results we have
calculated the different  
solutions of the Noble and two-variable models as a function of cable
length $L$ and pacing period $\tau$. As can be seen in 
Fig. \ref{phasediag} these results are in good agreement 
over a wide range of $L$ and $\tau$ with those obtained simulating 
the amplitude equations. In particular, discordant alternans appear as
soon as the cable is long enough to accommodate roughly a fourth of a wavelength,
as given by expressions (\ref{lsta}) and (\ref{ltra}) (see Table
\ref{lengths}). Several comments are in order regarding our results. 

(i) The transition from 
concordant to discordant alternans occurs as the length of the tissue 
is increased,
but the transition line remains nearly constant with increasing pacing 
rate. 
This seems to contradict the results obtained in 
\cite{Pasetal}, where a transition from concordant to discordant alternans 
was
reported as the pacing rate was increased. In typical models, the 
slope of CV-restitution is small for large values of DI, 
but increases very 
rapidly for small values of DI (see, for instance, the 
CV-restitution curve of the Noble model in Fig. \ref{fig.rest}b). Thus, 
as the pacing rate is 
increased, the amplitude of alternans grows thereby
engaging the steep part of the CV-restitution curve and
causing a transition to discordant alternans. 
This does not occur in the Noble model since 
there is a reverse period-doubling bifurcation at larger pacing rates by which
the steady state regains stability, and the oscillations never get to grow 
enough, but we have checked 
this 
point with simulations of the Beeler-Reuter equations (not shown here).  
Within our amplitude equations this effect can not be captured
since we are expanding around the period-doubling bifurcation point,
where the slope of the CV-restitution is fixed. However, it can be 
observed with the coupled map
equations. 

(ii) Discordant alternans appear directly from the rest state. This was 
already observed in simulations of the Beeler-Reuter model \cite{Watetal}, 
for lengths of tissue 
$L \gtrsim 5$ cm. As can be seen in Fig. 5 of that paper, the number of 
nodes
increases with the size of the tissue. This is also the case in our 
simulations
of the Noble model (Fig. \ref{fig.sim}) where, close 
to 
onset, the oscillations of APD adopt a sinusoidal form, with a well 
defined 
wavelength. A similar result was observed 
in \cite{Quetal}, where the Luo-Rudy model was modified to obtain several 
APD
and CV-restitution curves. In agreement with our results, when the slope 
of
CV-restitution was large at onset, a direct transition to discordant
alternans 
was observed, while, in the case of a shallower curve, the primary 
transition 
was to concordant alternans, discordant alternans only appearing when the 
oscillations in APD grew enough to explore the steep part of the 
CV-restitution curve.

(iii) The onset of alternans in an extended tissue is delayed with respect 
to that of the single cell (as given by Eq. (\ref{res})). This
restabilizing effect is due to a combination of diffusive coupling effects and 
CV-restitution (see Eqs. (\ref{onset.Nob})-(\ref{onset.2var})), that induces
oscillations in the period of stimulation that
effectively act as a control mechanism. In Eq. (\ref{aeqf}) this effect is 
produced by the integral term. As is readily seen in Fig. \ref{phasediag}
from the diagrams of the Noble and two-variable models, the
larger the slope of CV-restitution is at the critical point $f'=1$ (the 
smaller $\Lambda$ is), the larger the value of the restabilization. 

(iv) The nodes separating regions oscillating out of phase may be 
stationary
(Noble model), or traveling (two-variable model). Since this distinction
occurs already at onset, and there is no transition from one case into 
the other as the pacing rate is increased, the selected type must depend 
on the specific ionic model considered. There is experimental
evidence for both types of patterns \cite{Pasetal,Foxetal}. 
In real tissue, however, other effects not considered here, such as
gradients of restitution properties, could be important, and affect
the motion of the node. In the experiments in \cite{Foxetal} the tissue was
paced from both ends, with identical results, which seems to suggest
that gradients are not important in that case. In other situations,
however, the gradient terms could 
become important, and create a drift that opposes or reinforces the
effect of dispersion.

\subsection{Nonlinear state: conduction blocks and front solutions}

Nonlinearities are mainly responsible for saturating the amplitude of
the linear modes, although both the wavelength and evolution of the 
pattern generally vary with distance from onset.
Furthermore, at high
amplitudes conduction blocks can be produced, which are the main
mechanism for the induction of reentry. Recently, systems of coupled
maps similar to Eqs. (\ref{res}) and (\ref{disp}) have been used
to study the appearance of conduction blocks \cite{FoGi02,HeRa05,CoVi05}. 
They are often produced right after the first
node in the APD oscillations \cite{FoGi02}, thus stressing the 
importance of discordant alternans for the
induction of reentry. Furthermore, the position of 
the wave block has been shown to vary if the system presents 
traveling nodes \cite{HeRa05}. 

Although strictly valid only close to onset, we will use
our amplitude equations to get an idea on where conduction blocks can be 
formed. 
A conduction block occurs wherever the APD is large enough, so ${\rm DI}=\tau -
{\rm APD} < {\rm DI}_{min}$, being ${\rm DI}_{min}$ the minimum diastolic interval
necessary for propagation. Let us first neglect the
diffusive coupling. Then, using Eq. (\ref{aeq}), which is the amplitude
equation corresponding to the maps (\ref{res}) and (\ref{disp}), it is
easy to show that the maximum value of the APD occurs just after the
discontinuity (see Fig. \ref{fig.bif}). Considering that the system has
reached steady state, and taking the spatial derivative of Eq. (\ref{aeq}), we
obtain $\partial_x b = (\sigma -3ga^2) \partial_x a$. Using that $\partial_x
b\simeq a/\Lambda$ to
eliminate $b$, then, from 
\be
\partial_x a=\frac{a}{\Lambda (\sigma - 3ga^2)}\label{eq.ampcb}
\ee
we obtain that the discontinuity occurs when $a^2=\sigma/3g$. At that
point $b=\sigma a - ga^3=-2\sigma^{3/2}/(3\sqrt{3g})$, if we take the
minus sign for $a$. As $b$ is continuous
through the discontinuity of APD (Fig. \ref{fig.bif}), the value of
$a$ after the discontinuity can be calculated from
\be
g a^3 -\sigma a + b=0,
\ee
which gives $a=2\sqrt{\sigma/3g}$. Then, the conduction block
occurs at a value of the period implicitly given by
\be
{\rm DI}_{min}=\tau-A_c- 2\sqrt{\sigma/3g},\;\;
{\rm with}\;\;\sigma=\frac{1}{2}f''(\tau - \tau_c),  
\ee
and at a point in space given by the position of the first
singularity, or
\be
x_0=\frac{1}{2}\Lambda (1-\ln \sqrt{3})f''(\tau-\tau_c),\label{xsing}
\ee
obtained solving Eq. (\ref{eq.ampcb}) with the initial condition
$a(0)=\sqrt{\sigma/g}$.

One should be very careful when considering this latter result, since the
addition of diffusive effects changes the position of the node, and
can make it travel. The reason is that,
without diffusion, arbitrarily steep gradients can develop, that
prevent the node from moving. Once diffusive effects are included,
this is no longer the case, and the node travels, unless its motion is
balanced by the drift term. In fact, there is a close analogy between
the amplitude equation (\ref{aeqf}) and the real Ginzburg-Landau
equation that has been extensively studied in the context of phase
transitions and front propagation \cite{CroHoh}. The dynamics is
richer here because the integral term originating from dispersion 
causes a non-local interaction of the fronts separating two
out-of-phase oscillating regions with the pacing end of the
cable. 

This is easier to see assuming that both dispersion and the drift terms are
small. Then, to first order, one recovers the Ginzburg-Landau equation
\begin{equation}
 \tau \partial_t a = \sigma a+\xi^2 \partial^2_x a -g a^3, 
\end{equation}
that has a stationary solution in the form of a front 
$a_0(x)=\sqrt{\sigma/g}\tanh[\sqrt{\sigma/(2\xi^2)}(x-x_0)]$, connecting at 
$x=x_0$
regions oscillating 
out of phase. Then, to calculate the correction due to dispersion and
asymmetric coupling, we can set $a(x,t)=a_0(x,x_0(t))+a_1(x,t)$, which, assuming 
$a_1$
small, results to first order in
\begin{equation}
(\sigma +\xi^2 \partial^2_x - 3g a_0^2)a_1=\tau \partial_t a_0 + w\partial_x a_0 
+
\frac{1}{\Lambda}\int^{x}_{0} a_0 (x') dx'. 
\end{equation}
In order for the system to have a solution, the right
hand side must be orthogonal to the left eigenvector 
of the linear operator, which in this
case is simply $\partial_x a_0$ since the operator
is self-adjoint. Then, we obtain
\begin{equation}
-\tau \frac{d x_0}{dt} \int^{L}_{0} [\partial_x a_0(x')]^2 dx' + w \int^{L}_{0}
[\partial_x a_0(x')]^2 dx' + \frac{1}{\Lambda} \int^{L}_{0}\partial_x a_0(x')
\left [
\int^{x'}_{0} a_0 (x'') dx''\right ]dx' =0. 
\end{equation}
Evaluating the integrals, it is found that the motion of
the node is given by
\be
\tau\frac{dx_0}{dt}=
w - \frac{3}{\Lambda}\sqrt{\frac{\xi^2}{2\sigma}}x_0.
\ee
For arbitrary dispersion the expression for the motion becomes more
complicated, but the effect is essentially the same.
The gradient term corresponding to $w$ makes the node 
move with constant velocity away from the pacing point, while the
integral term provides a ramp that makes one 
phase of oscillation
preferred over the other. When dispersion is weak, there is a point at
which the two effects balance each other and the node relaxes
to the  equilibrium
position  
\be
x_0^{eq}=\frac{w \Lambda}{3}\sqrt{\frac{2\sigma}{\xi^2}},
\ee 
which differs markedly from the prediction that neglects
diffusive coupling given by Eq. (\ref{xsing}).

\subsection{Two-dimensional paced tissue}

Let us now generalize our formalism to the case of a square piece of
tissue paced at one corner. The equivalent of Eq. (\ref{aeqb}) is simply
\be 
 \tau \partial_t a = \sigma a -g a^3  - b
- w \, (\hat{\bf n}\cdot \nabla) a+\xi^2 \nabla^2 a,\label{aeqb2d}
\ee
where $\hat{\bf n}$ is a unit vector normal to the direction of
propagation of the wavefront, and we
impose non-flux boundary conditions on the boundaries of the tissue.

\begin{figure}[p]
\centerline{
\includegraphics[width=16cm]{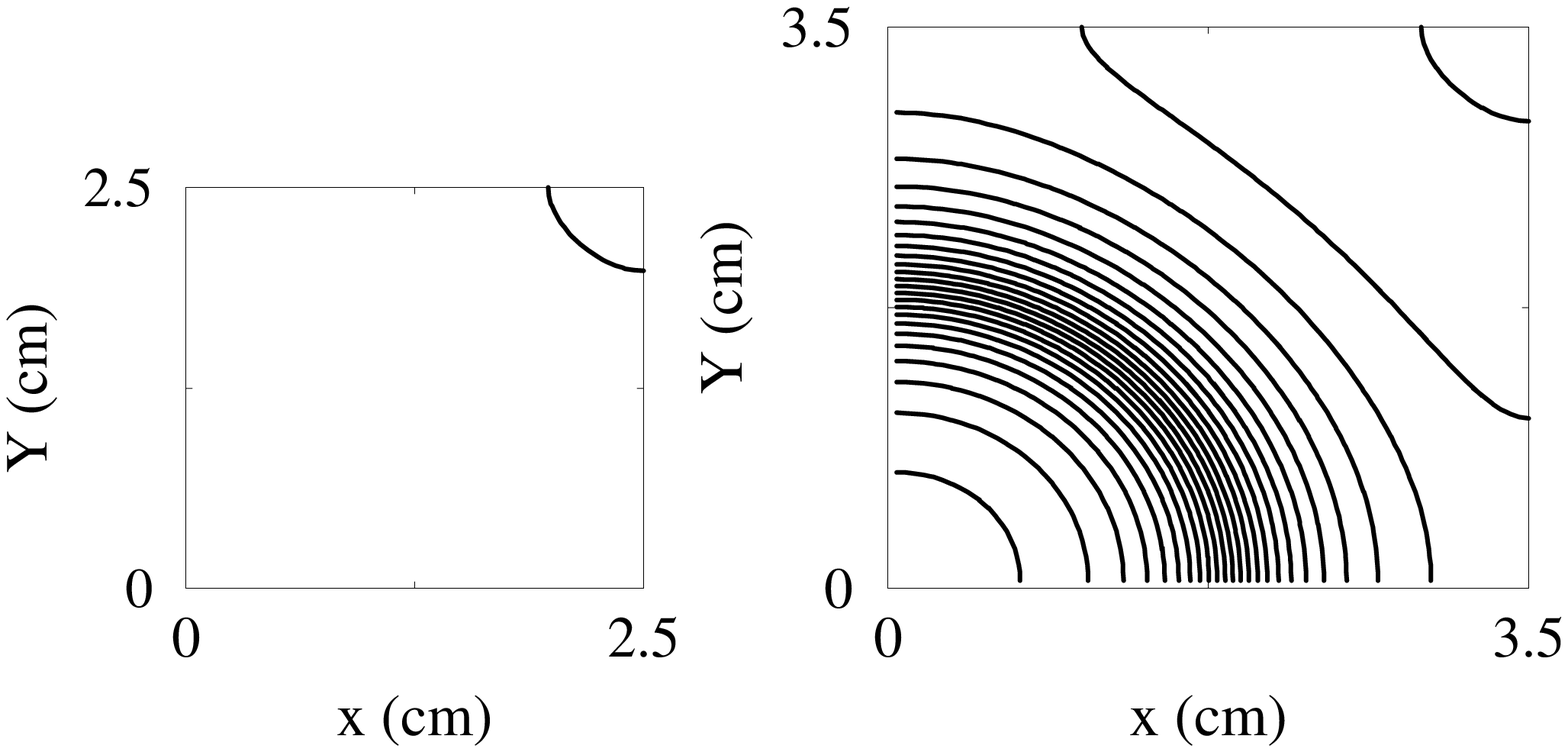}
}
\caption{Simulations of Eq. (\ref{aeqf2d}) for the parameters of
the Noble model, with  $\tau=258$ ms and a) $L=2.5$ cm, b) $L=3.5$ cm. The
solid lines represent the position of the node. The lines 
in b) are drawn every ten beats. As the size of the tissue increases, a 
node
forms at the corner opposite the pacing point (a), which above a certain 
tissue size begins to travel, due to curvature effects (b).
\label{fig.2Dnodes}}
\end{figure}

Now the determination of the period at a given point would involve the
integral along the path traveled by the wavefront. In particular, we
obtain the condition for the oscillations in period:
\be
(\hat{\bf n}\cdot \nabla)b=\frac{a}{\Lambda}. 
\ee
In general this is a complicated problem, since we have to know the  
normal to the wavefront at any given point.  A major
simplification occurs if we assume that the oscillations in APD do
not affect the form of the propagating wavefront, which is the same as
to say that the azimuthal variations of APD are small compared with
the radial ones. Then, as long as the thickness of the wavefront is
small compared with the size of the tissue, it can be assumed that,
except for narrow boundary layers at the borders of the tissue, the
propagation of the wavefront is perfectly circular. In this case, Eq. 
(\ref{aeqb2d}) can be rewritten as:
\be 
 \tau \partial_t a = \sigma a -g a^3  - \int^r_0 a dr'
- w \, \partial_r a+\xi^2 \nabla^2 a.\label{aeqf2d}
\ee
In the far field limit, these equations reduce to Eq. (\ref{aeqf}), 
with the node now becoming a circular line. We therefore expect to be
a minimum tissue size for the node to form (Fig. \ref{fig.2Dnodes}a),
as in the one-dimensional case. This was already observed in
experiments \cite{Pasetal} and numerical simulations of ionic models
\cite{Watetal}, where after a few beats, a nodal line was formed at
the corner opposite the pacing point, and then moved towards it. Close
to the pacing point, however, the motion of the node  becomes
nontrivial, because of curvature effects. Besides the term
$-w\partial_r a$ that tends to make the nodal line move in the radial
direction away from the pacing point, there is a contribution to the drift
in the direction normal to the interface coming from the Laplacian,
this being positive or negative depending on the curvature of the
interface. As the nodal line is always perpendicular to the
boundaries, its curvature depends on its position on the square. When
the tissue is large enough so in the one-dimensional case the node
would form in the region with positive curvature (close to the pacing
point), this extra effect may be enough to make it travel
(Fig. \ref{fig.2Dnodes}).

\section{Discussion\label{sec.disc}}

T-wave alternans is thought to play a key role in the 
transition from normal heart rhythm to reentrant
tachycardia and fibrillation. Alternans has been 
shown to induce spiral break-up \cite{Kar94,Co96},
leading to a disordered spatio-temporal state that is usually identified 
with fibrillation. For this reason, it has been hypothesized that a 
steep restitution 
curve may be a possible determinant of the transition from
VT to VF \cite{Kar94,Co96}. 
Although some experiments support this hypothesis \cite{Fib}, 
there is so far no conclusive evidence \cite{GiOt97}. From a theoretical 
point of view, and despite
earlier progress \cite{Kar94}, we still do not have a complete
understanding on how  
alternans affects the stability of spirals. Numerical simulations show
that break-up occurs when the amplitude of alternans oscillation  
grows enough to induce conduction blocks, which typically happens away from
the spiral core. However, there is presently no theory that predicts the spatial
distribution of alternans in a spiral wave. Topological arguments impose
the existence of a nodal line (also termed line defect) 
\cite{GoKa96}, with a jump in $2\pi$ in 
the phase of the oscillation, stretching from the core of the spiral to the 
boundaries. Thus, discordant alternans is always present, but what determines 
the motion of the nodal line, and whether it is important for spiral 
break-up is not entirely clear, except far from the spiral core where
the dynamics is essentially one-dimensional.     

In the present paper, we have considered the simpler cases of one and
two-dimensional paced tissue and a circulating pulse in a ring geometry, 
and derived an equation for the spatio-temporal dynamics of small amplitude 
alternans in these states. 
Although conduction blocks fall out of the scope
of the present theory, since it only considers small variations in 
conduction velocity, the results do shed light
on where conduction blocks will occur. For
that one can look at the distribution of APD and locate the places where
it is larger than a critical value. To study the evolution after a
conduction
block has occurred, one has to resort again to simulations of the original
equations. 

A circulating pulse in a ring has long been considered as a 
simplified model of anatomical reentry \cite{Mi14}. For rings smaller
than a critical value, oscillations in APD appear \cite{FraSim88}. An 
understanding of these oscillations can be helpful in terminating
anatomical reentry circuits. In this case discordant alternans is always
present, and its wavelength is 
determined by the length of the ring, with the fastest growing mode
having $\lambda \simeq 2L$. 

In the paced case, it has been shown \cite{Pasetal} that the gradients 
of repolarization created during discordant alternans offer a substrate 
for conduction block, leading to reentry and ventricular fibrillation. 
We have shown here that discordant wave patterns 
\cite{Pasetal,Foxetal,Quetal,Watetal} result from a finite wavelength
linear instability. Hence, their formation requires a minimum tissue size 
$L_{min}\sim \lambda/4$, required for at least
one node to form. The value of $L_{min}$ that we
measure in simulations of reaction-diffusion
models are actually close to $\lambda/4$
with $\lambda$ predicted by Eq. (\ref{lsta})
and Eq. (\ref{ltra}), respectively (Table \ref{lengths}). This lengthscale
is similar in two-dimensional paced tissue, although in this case, 
curvature
of the nodal line may affect the motion of this line. In both paced
one- and two-dimensional tissue and the ring geometry, 
the onset of alternans in tissue is different
than in a paced isolated cell because alternans, in tissue,
is manifested as a wave and diffusive coupling tends to smooth out 
spatial gradients of APD. 

The paced and reentrant geometries studied 
in the present paper also form the basis to 
develop a theory of the 
dynamics of alternans in the presence
of spiral waves, which has been the subject
of both experimental and numerical studies \cite{GoKa96}. 
Far from the core, the spiral is similar to the two-dimensional paced 
case. 
In the case where nodes move towards the pacing site in a one-dimensional 
cable, 
the nodal line should form a spiral with an opposite chirality as
the propagating spiral wave front. This opposite chirality is imposed
by the fact that the nodal line moves inward towards the core under
the effect of a sufficiently steep, positively sloped, 
CV-restitution curve, and under
the assumption that cellular 
alternans is driven by APD restitution. 
The wavelength of the nodal line
spiral far from the spiral core should be equal to the wavelength of 
discordant alternans in the one-dimensional paced case. This 
should
match to the nodal line expected from one-dimensional reentry close to the
spiral core. Then, the nodal line should extend straight out of the spiral core
in the simplest limit of a constant wave speed where the wavelength of
discordant
alternans diverges far from the core. 
A theory for the spiral must, therefore, reduce to the two cases studied 
in
this paper in the appropriate limits.
The development of a theory of alternans in this case remains a 
fascinating
task for future work.

\section{Conclusions\label{sec.conc}}

We have derived an equation that describes the spatio-temporal dynamics 
of small oscillation alternans in cardiac tissue. Our formulation is
based on the restitution properties of the system, and takes also into 
account intercellular coupling, which is crucial in order to derive 
the correct threshold pacing rate 
and lengthscale of discordant alternans. For a simplified 
two-variable ionic model, we have been able to calculate
these coefficients, and show that our reduced description agrees well with
numerical simulations of the model. We have also considered the more 
realistic Noble model, and measured these coefficients numerically, 
stressing the generality of our formulation. We have applied our
formulation 
to two different cases, a paced tissue and a circulating pulse in a ring.

In the ring, the amplitude equation predicts that both dispersion and 
intercellular coupling affect the frequency of the motion, and the 
oscillations of APD are typically quasiperiodic. There is
a particular value of the length of the ring where these two effects
balance
each other, and the oscillations become periodic. 
For the paced case, the amplitude equation predicts
that discordant wave patterns result
from a finite wavelength 
linear instability. Hence, their
formation requires a minimum tissue size 
$L_{min}\sim \lambda/4$, required for at least one node to form. 
This lengthscale is intrinsic, in opposition to the
circulating pulse, where it depends on the length of the ring. 
From the linear problem we deduce the possibility of two different patterns,
depending on the parameters of the system, a standing wave pattern, or a 
traveling one. In the latter state, the amplitude of the oscillation
grows exponentially away from the pacing point in a linear regime, 
which favors the induction of conduction blocks. 

Our formulation is based on the assumption that the simple map relationship
(\ref{res}) is satisfied. The addition of memory effects would modify the coefficients of the amplitude
equation, but not its general form, that is generic close to the period
doubling bifurcation point. This modification, however, has important 
consequences for relating the genesis  
of discordant alternans to the underlying cell physiology. As will be shown
elsewhere, one interesting
effect of memory is that it makes the coefficient
 of the non-local term in Eq. (\ref{eq.aintro}) depend both on the slope
 of the CV-restitution curve and the beat-to-beat dynamics. This opens
 the possibility to prevent the formation of discordant alternans by
 modifying the dynamics at a single-cell level to make 
 this coefficient negative, as would be the case for a
 negatively sloped ``supranormal'' CV-restitution curve when the
 single cell dynamics is governed simply by
  APD-restitution.

Even though the present theory does not take into account several effects
present in real tissue (such as the three dimensional 
fiber geometry of an anatomical heart, the corresponding 
anisotropy in conduction velocity, spatial gradients of restitution 
properties, etc) it sheds light on the basic mechanisms and length scales
that control the formation and evolution of discordant alternans.
Moreover, it identifies a small subset of 
relevant parameters that control the formation of these 
arrhythmogenic patterns in experiments and numerical 
simulations of more complex ionic models. 

\acknowledgments

This work was supported by National Institutes of Health/National
Heart, Lung, and Blood Institute Grants P50-HL52319 and P01 HL078931.  
B.E. wants to acknowledge financial support by MCyT (Spain), and by MEC
(Spain), under project FIS2005-06912-C02-01.

\appendix

\section{Action potential duration and conduction velocity of the two-variable
  model\label{ap2var}}

In this appendix we show how to obtain the APD and CV-restitution curves for a
propagating pulse in the two variable model. We will consider the limit
$\epsilon \rightarrow 0$ in which the sigmoidal becomes a step
function $S(V-V_c)\rightarrow \Theta(V-V_c)$. For a pulse propagating
with velocity $c$ in the right direction, we can write
$\partial_t=-c\partial_x$, and Eqs. (\ref{eq.I2var})-(\ref{eq.h2var})
become:
\begin{eqnarray}
\left \{
\begin{array}{c}
D\partial^2_x V+c\partial_x V- 1/\tau_0 +h/\tau_a=0\\
-c\partial_x h=-h/\tau_{+}
\end{array}
\right . & \;{\rm when} & \; V> V_c,\\
\left \{
\begin{array}{c}
D\partial^2_x V+c\partial_x V- V/(V_c\tau_0) =0\\
-c\partial_x h=(1-h)/\tau_{-}
\end{array}
\right . & \;{\rm when} & \; V <  V_c,\\
\end{eqnarray}
and we will assume that, at $x=0$ we have the position of the wavefront given
by $V(0)=V_c$.

The equations for the gate $h$ can be integrated to give:
\begin{eqnarray}
&&h(x)=h_0 e^{x/(c\tau_{+})},\; {\rm when} \; V> V_c,\label{eq.apph1}\\ 
&&h(x)=1-h_1 e^{x/(c\tau_{-})},\;{\rm when}\; V < V_c,\label{eq.apph2}
\end{eqnarray}
where $h_0$ and $h_1$ are constants of integration. If we assume that ${\rm APD} >>
\tau_{+}$ (${\rm APD} \sim 200$ ms, and $\tau_{+}=12$ ms in our simulations), then
we can consider that $h$ has decayed to zero by the end of the action
potential, i.e., at the waveback of the previous pulse. Then $h(c {\rm DI})=0$, and
using Eq. (\ref{eq.apph2}) we obtain $h_1=\exp{(-{\rm DI}/\tau_{-})}$, and
$h_0=h(0)=1-\exp{(-{\rm DI}/\tau_{-})}$.
Next, we can solve the equation for the potential that, when $V < V_c$ becomes 
\begin{equation}
V(x)=Ae^{\lambda_1 x}+Be^{\lambda_2 x},
\end{equation}
with 
\begin{equation}
\lambda_{1,2}=\frac{1}{2D}\left [-c\mp\sqrt{c^2+4D/(\tau_0 V_c)}\right ].
\end{equation}  
A solution in the wavefront ($x>0$) must satisfy that $V \rightarrow 0$
when $x$ is large and positive. Therefore it will correspond to a decaying
mode:
\begin{equation}
V(x\geq 0)=V_c e^{\lambda_1 x}=V_c \exp{\left 
[\frac{x}{2D}(-c-\sqrt{c^2+4D/(\tau_0 V_c)})\right
  ]}. 
\end{equation}
In the waveback ($x<-{\rm APD} c$), on the other hand, we must impose that $V
\rightarrow 0$ 
when $x$ is large and negative, corresponding to a growing mode in
space. Then:
\begin{equation}
V(x\leq -{\rm APD} c)=V_c e^{\lambda_2 (x+{\rm APD} c)}=V_c \exp{\left [\frac{x+{\rm
        APD} 
c}{2D}(-c+\sqrt{c^2+4D/(\tau_0
      V_c)})\right ]}. 
\end{equation}
The value of the potential for $-{\rm APD} c < x < 0$ will be given by the solution
of
\begin{equation}
D\partial^2_x V+c\partial_x V- 1/\tau_0 +h_0 e^{x/(c\tau_{+})}/\tau_a=0,
\end{equation} 
that can be written as:
\begin{equation}
V(x)=Ae^{-cx/D} +B +\frac{x}{c\tau_0} -\frac{c^2 \tau^2_{+}}{D+c^2
  \tau_{+}}\frac{h_0}{\tau_a}e^{x/c\tau_+}, \;{\rm for} \; -{\rm APD} c < x < 0, 
\end{equation}
where we recall that $h_0=1-\exp{(-{\rm DI}/\tau_{-})}$.
Then, if we obtain the constants $A$ and $B$ imposing continuity of the potential
and its derivative at $x=0$, the same conditions at $x=-{\rm APD} c$ will give us
the value of the APD and conduction velocity $c$. Imposing the bc's at $x=0$
and $x=-c {\rm APD}$ we obtain the set of equations:
\begin{eqnarray}
&&V_c=A+B -\frac{c^2 \tau^2_{+}}{D+c^2
  \tau_{+}}\frac{h_0}{\tau_a},\\
&&V_c=-\frac{{\rm APD}}{\tau_0} + Ae^{c^2 {\rm APD}/D} + B -\frac{c^2 \tau^2_{+}}{D+c^2
  \tau_{+}}\frac{h_0}{\tau_a}e^{-{\rm APD}/\tau_{+}},\\
&&V_c \lambda_1=-\frac{c}{D}A + \frac{1}{c\tau_0} -\frac{c \tau_{+}}{D+c^2
  \tau_{+}}\frac{h_0}{\tau_a},\\
&&V_c \lambda_2= -\frac{c}{D}Ae^{c^2 {\rm APD}/D} + \frac{1}{c\tau_0} -\frac{c 
\tau_{+}}{D+c^2
  \tau_{+}}\frac{h_0}{\tau_a}e^{-{\rm APD}/\tau_{+}}. 
\end{eqnarray}
We have four equations for the four unknowns $A$, $B$, ${\rm APD}$ and
$c$. These equations can be simplified provided that ${\rm APD} \gg \tau_{+}$ and
$D/c \ll c {\rm APD}$. Then, we will define a new constant $C=Ae^{c^2 {\rm APD}/D}$, and
take the approximations $\exp(-c^2 {\rm APD}/D) \simeq 0$, $\exp(-{\rm APD}/\tau_{+}) \simeq
0$, from which we obtain
\begin{eqnarray}
&&V_c=B -\frac{c^2 \tau^2_{+}}{D+c^2
  \tau_{+}}\frac{h_0}{\tau_a},\label{eq.bcs1}\\
&&V_c=-\frac{{\rm APD}}{\tau_0} + C + B,\label{eq.bcs2}\\
&&V_c \lambda_1=\frac{1}{c\tau_0} -\frac{c \tau_{+}}{D+c^2
  \tau_{+}}\frac{h_0}{\tau_a},\label{eq.bcs3}\\
&&V_c \lambda_2= -\frac{c}{D}C + \frac{1}{c\tau_0}. \label{eq.bcs4}
\end{eqnarray}
Eq. (\ref{eq.bcs3}) directly gives an implicit relation for the conduction 
velocity $c$
in terms of the diastolic interval DI:
\begin{equation}
\frac{c\tau_{+}}{D+c^2\tau_{+}}\frac{1-e^{-{\rm DI}/\tau_{-}}}{\tau_a}=\frac{c
  V_c}{2D}\left (1+\sqrt{1+\frac{4D}{c^2\tau_0 V_c}}\right ) +
\frac{1}{c\tau_0}.\label{restCVcable}
\end{equation}
The value of the APD can be obtained from Eq. (\ref{eq.bcs2}):
\begin{equation}
{\rm APD}=\tau_0 (B+C-V_c),
\end{equation}
where 
\begin{equation}
B=V_c+\frac{c^2 \tau^2_{+}}{D+c^2
  \tau_{+}}\frac{h_0}{\tau_a}.
\end{equation}
To obtain the constant $C$ instead of directly using Eq. (\ref{eq.bcs4}) it is
more convenient to take the difference of Eqs. (\ref{eq.bcs3}) and
(\ref{eq.bcs4}) from which we obtain
\begin{equation}
C=\frac{D V_c}{c}(\lambda_1-\lambda_2)+ \frac{D \tau_{+}}{D+c^2
  \tau_{+}}\frac{h_0}{\tau_a}
\end{equation} 
Then, the APD becomes:
\begin{equation}
{\rm APD}=\tau_0 \left [\frac{\tau_{+}}{\tau_a}(1-e^{-{\rm DI}/\tau_{-}})-V_c
\sqrt{1+\frac{4D}{c^2 \tau_0 V_c}}\right ].
\end{equation}
To calculate the APD one has to solve then first Eq. (\ref{restCVcable}) for
the conduction velocity. This can be avoided neglecting the term $4D/(c^2 \tau_0 
V_c)$ in the
square root, that introduces an error of 1 or 2 ms in the determination of the
APD. This is equivalent to neglecting the effect of the outward repolarization
current $1/\tau_0$ during depolarization. In this way, CV and APD-restitution 
become
decoupled, and we can write the final expression 
\begin{equation}
{\rm APD}=\frac{\tau_{+}\tau_0}{\tau_a}(1-e^{-{\rm DI}/\tau_{-}})-V_c 
\tau_0.\label{ec.restAPDcable}
\end{equation}

\section{Derivation of the kernel\label{appA}}

Let us show how to calculate the kernel in 
Eq. (\ref{eq.kernel}), and
the coefficients $w$ and $\xi^2$, for the two-variable model.
This will allow us to verify the theory, as we can compare our 
analytical results with the simulations, and also gain some physical 
intuition on the origin of these coefficients. 
To derive the kernel, we start from
the cable equation (\ref{cable}) (with $I_{ext}=0$). Our main goal is to 
quantify the effect of a gradient of APD on the 
restitution curve. For that, we will consider a single pulse 
propagating in this gradient with speed $c$. Interpreting the cable 
equation as
a diffusion equation where $I_{ion}$ is a spatially distributed
source, Eq. (\ref{cable}) can be formally inverted, expressing the 
transmembrane 
potential $V$ in 
terms of the Green function of the diffusion operator:
\be
V(x,t)=\frac{1}{C_m}\int^t_{-\infty} dt' \int^{\infty}_{-\infty} dx'
\frac{\exp\left \{-\frac{(x-x')^2}{4D(t-t')}\right \}}
{[4\pi D(t-t')]^{1/2}} I_{ion}(x',t'),
\label{eq.invertV}
\ee
where we have assumed translational invariance, as is the case in the
ring. 

Now, if we are measuring the action potential duration at a given
value of the transmembrane potential $V_c$, then, by definition
\be
V_c=V(x,x/c+A(x)),
\ee
where $x/c$ is the time it took the pulse to arrive at position $x$ and,
for simplicity we introduce the notation $A(x)\equiv {\rm APD}(x)$. 
Substituting this into Eq. (\ref{eq.invertV}), we obtain
\be
V_c=\frac{1}{C_m}\int^{x/c+A(x)}_{-\infty} dt' \int^{\infty}_{-\infty} dx'
\frac{\exp\left \{-\frac{(x-x')^2}{4D(x/c+A(x)-t')}\right \}}
{[4\pi D(x/c+A(x)-t')]^{1/2}}I_{ion}(x',t').
\label{eq.vc}
\ee
This defines an implicit integral equation for the action potential duration.

In the general case it is not clear how to
invert this equation. We can do it, however, for the simpler case of
the two variable model, where the action potential adopts a triangular
form. Integrating Eq. (\ref{eq.h2var}) for the gate variable $h$ (in the limit
$\epsilon \rightarrow 0$), for $V> V_c$, we obtain $h=h_0 e^{-t/\tau_+}$,
where $h_0=1-e^{-{\rm DI}/\tau_-}$ can be obtained integrating Eq. (\ref{eq.h2var}) for 
$V<V_c$, and
assuming $\tau_+ \ll {\rm APD}$, so by the end of the previous action potential the 
gate is
completely closed ($h=0$).

Then, when $V>V_c$, and in the absence of propagation, the equation for the 
temporal evolution of the 
transmembrane potential becomes:
\be
\dot{V}=\frac{1}{\tau_a}(1-e^{-{\rm DI}/\tau_{-}})e^{-t/\tau_{+}} 
- \frac{1}{\tau_0}.\label{eq.dVdt}
\ee
When $\tau_{+}\ll\tau_0 $ we can assume that depolarization occurs 
instantaneously, so we can write the current in the form
\be
I_{ion}(t)/C_m=-\theta(t) I+ \theta(t-{\rm APD}) I + \delta(t)J[{\rm DI}],
\label{ap.Iion}
\ee
where $\theta(t)$ is the standard
Heaviside step function, and $\delta(t)$ the Dirac delta 
function. The former equation means that, after an excitation at $t=0$, the 
voltage takes the maximum value $V_{max}=J[{\rm DI}]$ and then 
decreases linearly in time $V=V_{max}-It$, with the  
identifications $J[{\rm DI}]=\tau_{+}(1-e^{-{\rm DI}/\tau_{-}})/\tau_a$ and
$I=1/\tau_0$. 

Integrating Eq. (\ref{eq.dVdt}), under the approximation $\tau_{+}\ll\tau_0 $ we 
obtain
\be
V(t)=\frac{\tau_{+}}{\tau_a}(1-e^{-{\rm DI}/\tau_{-}}) 
- \frac{t}{\tau_0}=V_{max} - \frac{t}{\tau_0}.   
\ee
At $t={\rm APD}$, $V({\rm APD})=V_c$, and the value of the
APD becomes ${\rm APD}=(V_{max}-V_c)\tau_0$, or
\be
{\rm APD}^{n+1}=(J[{\rm DI}^n]-V_c)/I=\frac{\tau_{+}
  \tau_{0}}{\tau_{a}}(1-e^{-{\rm DI}/\tau_{-}})-V_c \tau_0 \equiv f({\rm DI}^n),
\ee
which gives us the APD-restitution. This is the same expression that we
obtained in the previous appendix (cf. Eq. (\ref{ec.restAPDcable})),
neglecting the term $4D/(c^2 \tau_0 V_c)$.

For a propagating pulse in a gradient of DI, we can again use expression
(\ref{ap.Iion}), but taking into account that the excitation now occurs
when the pulse arrives (i.e. at time $t=x/c$), and that the diastolic interval
depends on space. Then
\be
I_{ion}(x,t)/C_m=-\theta(t-x/c) I+ \theta (t-x/c-A(x)) I +  \delta(t-x/c)J[{\rm DI}(x)].
\label{ap.Iionx}
\ee
It is important to emphasize that $V_{max}=J[{\rm DI}]$ is different for
an isolated cell than for a cable because of diffusive coupling. Therefore
$J[{\rm DI}]$ in Eq. (\ref{ap.Iionx}) denotes the peak value of the voltage
for a propagated action potential, as opposed to a stimulated cell 
in Eq. (\ref{ap.Iion}). In contrast, the local repolarizing current $I$
is the same in both cases. 

Substituting this expression into (\ref{eq.vc}) we can now split the
integral into three parts $V_c=V_I+V_{II}+V_{III}$. The first part is
\bea
V_I&=&\int^{x/c+A(x)}_{-\infty} dt' \int^{\infty}_{-\infty} dx' 
\frac{\exp\left \{-\frac{(x-x')^2}{4D(x/c+A(x)-t')}\right \}}
{[4\pi D(x/c+A(x)-t')]^{1/2}}
\theta(t'-x'/c) I\nonumber \\
&=&I \int^{x/c+A(x)}_{-\infty} dt' \int^{ct'}_{-\infty} dx'
\frac{\exp\left \{-\frac{(x-x')^2}{4D(x/c+A(x)-t')}\right \}}
{[4\pi D(x/c+A(x)-t')]^{1/2}}.
\eea
This gives
\be
V_I=I \int^{x/c+A(x)}_{-\infty} dt' \frac{1}{2}
\left \{1+Erf \left [\frac{ct'-x}{\sqrt{4D(x/c+A(x)-t')}} \right ] \right \}.
\ee
Making the approximation 
\be
\frac{1}{2}
\left \{1+Erf\left [\frac{y}{\sqrt{4D(A(x)-y/c)}}\right ] \right \}
\simeq \theta(y),
\ee 
valid when $D/c \ll (DA)^{1/2} \ll Ac$, we obtain
\be
V_I\simeq I \int^{x/c+A(x)}_{-\infty} dt' \theta(ct'-x) =IA(x).\label{iax}
\ee 
As long as the length scales associated with intercellular 
coupling are small compared with the distance the pulse travels during an
APD (so the coupling with other cells is negligible), this integral gives us
the change in voltage during repolarization at a given point, during
the time of one APD. 
   
The second part becomes:
\be
V_{II}=\int^{x/c+A(x)}_{-\infty} dt' \int^{\infty}_{-\infty} dx' 
\frac{\exp\left \{-\frac{(x-x')^2}{4D(x/c+A(x)-t')}\right \}}
{[4\pi D(x/c+A(x)-t')]^{1/2}}\theta(t'-x'/c-A(x')) I 
\ee
To be able to solve for $V_{II}$ we make the approximation $A(x) \simeq A_c$. Then:
\bea
V_{II} &\simeq& I \int^{x/c+A_c}_{-\infty} dt' \int^{c(t'-A_c)}_{-\infty} dx'
\frac{\exp\left \{-\frac{(x-x')^2}{4D(x/c+A_c-t')}\right \}}
{[4\pi D(x/c+A_c-t')]^{1/2}}\nonumber \\
&=&I \int^{x/c+A_c}_{-\infty} dt' \frac{1}{2}
\left \{1+Erf \left [-\frac{c}{\sqrt{4D}} \sqrt{x/c+A_c-t'} \right ] \right \}.
\eea
Making the change of variable $y=c^2 (x/c+A_c -t')/(4D)$ we obtain:
\be
V_{II} \simeq \frac{2 D}{c^2} I \int^{\infty}_{0} dy \left [ 1 +
  Erf(-\sqrt{y}) \right ] = \frac{D}{c^2} I
\ee 
Clearly this term is much smaller that the previous one ($V_{II}/V_{I} \simeq
(D/c)/(c A_c) \ll 1$) and we will neglect it. 

The last term is 
\bea
V_{III}&=&\int^{x/c+A(x)}_{-\infty} dt' \int^{\infty}_{-\infty} dx'
\frac{\exp\left \{-\frac{(x-x')^2}{4D(x/c+A(x)-t')}\right \}}
{[4\pi D(x/c+A(x)-t')]^{1/2}} J[{\rm DI}(x')]\delta(t'-x'/c)\nonumber \\
&=&\int^{x/c+A(x)}_{-\infty} dt' 
\frac{\exp\left \{-\frac{(x-ct')^2}{4D(x/c+A(x)-t')}\right \}}
{[4\pi D(x/c+A(x)-t')]^{1/2}}cJ[{\rm DI}(ct')].
\eea
Making the change of variable $t'=(x+y)/c$, we finally obtain
\be
V_{III}=\int^{cA(x)}_{-\infty} dy 
\frac{\exp\left \{-\frac{y^2}{4D(A(x)-y/c)}\right \}}
{[4\pi D(A(x)-y/c)]^{1/2}}J[{\rm DI}(x+y)].
\ee
And this integral gives us the correction to the maximum value of the
APD coming from the intercellular coupling, as $V_{III}\simeq V_{max}$. 

Adding $V_{I}$ and $V_{III}$, and solving for $A(x)$ in Eq. (\ref{iax}), we obtain
\bea
A(x)&=&\int^{\infty}_{-\infty} dy 
\frac{\exp\left \{-\frac{y^2}{4D(A(x)-y/c)}\right \}}
{[4\pi D(A(x)-y/c)]^{1/2}}(J[{\rm DI}(x+y)]-V_c)/I \nonumber \\
&\equiv &\int^{\infty}_{-\infty} dy 
\frac{\exp\left \{-\frac{y^2}{4D(A(x)-y/c)}\right \}}
{[4\pi D(A(x)-y/c)]^{1/2}}f[{\rm DI}(x+y)],
\eea
where $A(x)$ is the APD following the diastolic interval DI, so 
$A(x)\equiv {\rm APD}^{n+1}$ and ${\rm DI}\equiv {\rm DI}^n$, and
we have extended the limits of the integral to infinity, assuming a
rapid decay of the kernel. Note that $A(x)$ is
still in the kernel, so the former equation gives an implicit
expression for the APD. As we are interested in the regime close to
the onset of period doubling, we can approximate the APD within the
integral as $A(x)\simeq A_c$. The deviations from the critical value
would give nonlinear gradient terms in the oscillations of APD, that
we assume to be of higher order. Then, our final expression is:
\be
{\rm APD}^{n+1}(x)= \int^{\infty}_{-\infty} dy 
\frac{\exp\left \{-\frac{y^2}{4D(A_c-y/c)}\right \}}
{[4\pi D(A_c-y/c)]^{1/2}}f[{\rm DI}^n(x+y)].
\ee
Once we know the expression for the kernel, the coefficients $w$ and $\xi^2$
are easy to obtain. It is useful to rewrite the kernel using the change of 
variable $y=2(D\,A_c)^{1/2}z$. Then, expanding the kernel to first
order in the coefficient $2(D\,A_c)^{1/2}/(A_c c)$, again assumed to
be small, we get 
\be
{\rm APD}^{n+1}(x)=\frac{1}{\sqrt{\pi}}\int_{-\infty}^{\infty}dz
e^{-z^2}[1+\frac{1}{c}\sqrt{\frac{D}{A_c}}(z-2z^3)]
f[{\rm DI}^n(x+2(D\,A_c)^{1/2}z)].\label{apd.expand}
\ee 
We can expand the restitution curve around its value at
a given point
\bea
f[{\rm DI}^n(x+y)]&=& f\left [{\rm DI}^n(x)+y\partial_x
{\rm DI}^n(x)+\frac{1}{2}y^2\partial^2_x {\rm DI}^n(x)+ \cdots \right ]\nonumber\\
&=&f[{\rm DI}^n(x)] + 2z(D\,A_c)^{1/2}f'\partial_x {\rm DI}^n(x) +
2z^2 D\,A_cf' \partial^2_x {\rm DI}^n(x)\nonumber \\
&&+ 2z^2 D\,A_cf'' [\partial_x {\rm DI}^n(x)]^2+\cdots
\eea
Introducing this expansion in Eq. (\ref{apd.expand}), the asymmetrical part results
\be
\frac{2D}{c}f'\partial_x 
{\rm DI}^n(x)\frac{1}{\sqrt{\pi}}\int_{-\infty}^{\infty}dz
e^{-z^2}(z^2-2z^4)=-\frac{2D}{c}f'\partial_x {\rm DI}^n(x),
\ee
and the symmetrical one
\be
2DA_c \left \{ f'\partial^2_x {\rm DI}^n(x) + f'' [\partial_x {\rm DI}^n(x)]^2 \right \}
\frac{1}{\sqrt{\pi}}\int_{-\infty}^{\infty}dz e^{-z^2}z^2=
DA_c \left \{f'\partial^2_x {\rm DI}^n(x) + f'' [\partial_x {\rm DI}^n(x)]^2 \right \}. 
\ee
In this last expresion we will neglect the term $[\partial_x {\rm DI}^n(x)]^2$ since,
close to the onset of alternans, it is higher order with respect to the term $\partial^2_x {\rm
  DI}^n(x)$  (their ratio is $\sim (f'/f'') \delta D$, where $\delta D$ is the
amplitude of the oscillations in ${\rm DI}(x)$). However, if the system is far from 
onset, it maybe necessary to take it into account, if we want to obtain 
quantitative results. 

Then, we have
\be
{\rm APD}^{n+1}(x)=f[{\rm DI}^n(x)]-\frac{2D}{c}f'\partial_x {\rm DI}^n(x)+
DA_c f'\partial^2_x {\rm DI}^n(x)+\cdots \label{devs}
\ee
This expression is the same as Eq. (\ref{expkernel}). Identifying
terms we obtain the values $w=2D/c$, $\xi^2=D\,{\rm APD}_c$. 
For more complicated ionic models these coefficients have to be 
calculated numerically following the procedure outlined in Sec. IV.A.

\newpage

\newpage

\newpage

\newpage

\newpage

\newpage

\pagebreak

\newpage

\newpage

\newpage

\newpage

\newpage

\newpage

\begin{thebibliography}{99}

\bibitem{Lewis} T. Lewis, Q. J. Med. {\bf 4}, 141 (1910).

\bibitem{precur} P.J. Schwartz, A. Malliani, Am. Heart J. {\bf 89}, 45 (1975);
J.M. Smith, 
E.A. Clancy, R. Valeri, J. N. Ruskin, R.J. Cohen,
Circulation {\bf 77}, 110 (1988);
B. Nearing, A. H. Huang, and R. L. Verrier, Science {\bf 252}, 437 (1991);
D. S. Rosenbaum, 
L. E. Jakson, J. M. Smith, H. Garam, J. N. Ruskin, R. J. Cohen,
N. Engl. J. Med. {\bf 330}, 235 (1994).

\bibitem{Pasetal}
J. M. Pastore, 
S.~D.~Girouard, K.~R.~Laurita, F.~G.~Akar and D.~S.~Rosenbaum, 
Circulation {\bf 99}, 1385 (1999).

\bibitem{Min} G. R. Mines, J. Physiol. (Lond.) {\bf 46}, 349 (1913).

\bibitem{Foxetal}
J. J. Fox, 
M. L. Riccio, F. Hua, E. Bodenschatz, and R. F. Gilmour,
Circ. Res. {\bf 90}, 289 (2002).

\bibitem{Quetal}
Z. Qu 
, A.~Garfinkel, P-S.~Chen, J.~W.~Weiss, 
Circulation {\bf 102},
1664 (2000).

\bibitem{Watetal}
M. A. Watanabe, 
F.~H.~Fenton, S.~J.~Evans, H.~M.~Hastings and A.~Karma, 
J. Cardiovasc. Electrophys. {\bf 12}, 196 (2001).  

\bibitem{Kar94} A. Karma, Chaos {\bf 4}, 461 (1994).

\bibitem{FeCh02}
F.F. Fenton, E.M. Cherry, H.M. Hastings and S.J. Evans, Chaos {\bf 12}, 852 (2002).

\bibitem{FraSim88} L.H. Frame and M.B. Simson, Circulation {\bf 78}, 1277 (1988).

\bibitem{Couetal}
M. Courtemanche, L. Glass, and J. P. Keener, 
Phys. Rev. Lett. {\bf 70}, 2182 (1993);
M.  Courtemanche, J. P. Keener, and L. Glass,
SIAM J. Appl. Math. {\bf 56}, 119 (1996). 

\bibitem{Vin}
A. Vinet, Annals Biomed. Eng. {\bf 28}, 704 (2000).

\bibitem{Ro01}
D. S. Rosenbaum, J. Cardiovasc. Electrophys. {\bf 12}, 207 (2001).

\bibitem{LaGi96}
K. R. Laurita, S. D. Girouard, and D. S. Rosenbaum, Circulation Research 
{\bf 79}, 493 (1996).

\bibitem{WeKa06}
J. N. Weiss, A. Karma, Y. Shiferaw, P. S. Chen, A. Garfinkel, Z. Qu, Circ. Res. 
\textbf{98}, 1244 (2006).

\bibitem{NolDal} J. B. Nolasco and R. W. Dahlen, J. App. Physiol, {\bf 25},
191 (1968).

\bibitem{Gueetal}
M. R. Guevara 
, G. Ward, A. Shrier, and L. Glass
in {\it Computers in Cardiology}, IEEE Comp. Soc., pp. 167 (1984). 

\bibitem{Fenton} F. H. Fenton, S. J. Evans, and H. M. Hastings, 
Phys. Rev. Lett. {\bf 83}, 3964 (1999).

\bibitem{HaBa99}
G.M. Hall, S. Bahar, and D.J. Gauthier, Phys. Rev. Lett. {\bf 82}, 2995 (1999).

\bibitem{FoBo02}
J.J. Fox, E. Bodenschatz, and R.F. Gilmour, Phys. Rev. Lett. {\bf 89},
138101 (2002).

\bibitem{ScCa04}
D. G. Schaeffer {\em et al}, Bull. Math. Bio. {\bf 69}, 459 (2007).

\bibitem{ToSc03}
E.G. Tolkacheva, D.G. Schaeffer, D.J. Gauthier, and W. Krassowska,
Phys. Rev. E {\bf 67}, 031904 (2003).

\bibitem{ShWa03}
Y. Shiferaw, M. Watanabe, A. Garfinkel, J. Weiss, and A. Karma,
Biophys. J. {\bf 85}, 3666 (2003).

\bibitem{ShSa05}
Y. Shiferaw, D. Sato, and A. Karma, Phys. Rev. E {\bf 71},
    021903 (2005).


\bibitem{EcKa02a} B. Echebarria and A. Karma, Phys. Rev. Lett. {\bf
88}, 208101 (2002). 

\bibitem{EcKa02b} B. Echebarria and A. Karma, Chaos {\bf 12}, 923 (2002).  

\bibitem{Chretal06} D. J. Christini, M. L. Riccio, C. A. Culianu, J. J. Fox, A. 
Karma, and R. F. Gilmour Jr., Phys. Rev. Lett. \textbf{96}, 104101 (2006).

\bibitem{CroHoh} M. C. Cross and P. C. Hohenberg,                       
Rev. Mod. Phys. {\bf 65}, 851 (1993). 

\bibitem{Nob}
D. Noble, J. Physiol {\bf 160}, 317 (1962). 

\bibitem{CytKee02}
E. Cytrynbaum and J. P. Keener,
Chaos \textbf{12}, 788 (2002).  



\bibitem{coeffs}
Eq. (\ref{aeqb}) becomes, in the case of the Noble model: 
$\tau \partial_t a =\sigma a - g a^3 -\chi a^5 - b -w\partial_x a
+\xi^2 \partial^2_x a$, with the coefficients
$\sigma=5.71\cdot 10^{-3} (\tau_c-\tau)$, $g\simeq -8\cdot 10^{-6}$,   
$\chi \simeq 1.37\cdot 10^{-8}$, and $\tau_c=262.5$ ms. For the two-variable
model we have used the  
analytical form of the APD-restitution curve, given by 
Eq. (\ref{eq.2var_APD}), 
so the coefficients are $\sigma= (\tau_c-\tau)/(2\tau_-) = 8.33\cdot 
10^{-3} 
(\tau_c-\tau)$, with $\tau_c=321.6$ ms, and $g=1/(12\tau_{-}^2)=2.31\cdot 
10^{-5}$.



\bibitem{Babetal} K. L. Babcock, G. Ahlers, and D. S. Cannell,
Phys. Rev. Lett. {\bf 67}, 3388 (1991).


\bibitem{FoGi02}
J.J. Fox, R.F. Gilmour, and E. Bodenschatz, Phys. Rev. Lett. {\bf 89},
198101-1 (2002).

\bibitem{HeRa05}
H. Henry and W.-J. Rappel, Phys. Rev. E {\bf 71}, 051911 (2005).

\bibitem{CoVi05}
P. Comtois, A. Vinet, and S. Nattel, Phys. Rev. E {\bf 72}, 031919 (2005).



\bibitem{Co96}
M. Courtemanche, Chaos {\bf 6}, 579 (1996);
J.N. Weiss, A. Garfinkel, H.S. Karaguezian, Z. Qu, and P.S. Chen,
Circulation {\bf 99}, 2819 (1999).

\bibitem{Fib} 
M. L. Riccio, M. L. Koller,
and R. F. Gilmour, Circ. Res. {\bf 84}, 955 (1999);
A. Garfinkel, 
Y. Kim, O. Voroshilovsky, Z. Qu, J.R. Kil, M-Y. Lee, H.S. Karaguezian, 
J.N. Weiss, and P-S. Chen,  Proc. Natl. Acad. Sci 
USA {\bf 97}, 6061 (2000).

\bibitem{GiOt97}
R.F. Gilmour, N.F. Otani, and M.A. Watanabe, Am. J. Physiol. {\bf 272},
H1826 (1997).

\bibitem{GoKa96}
A. Goryachev and R. Kapral, Phys. Rev. E {\bf 54}, 5469 (1996);
A. Goryachev, H. Chat\'e, and R. Kapral, Phys. Rev. Lett. {\bf 80}, 873 
(1998);
J-S. Park and K.J. Lee, Phys. Rev. Lett. {\bf 83}, 5393 (1999);
J.-S. Park, S. J. Woo, and K. J. Lee, Phys. Rev. Lett. {\bf 93},
098302 (2004); S. M. Hwang, T. Y. Kim, and K. J. Lee, Proc. Natl. Acad. 
Sci. USA {\bf 102}, 10363 (2005).

\bibitem{Mi14}
G.R. Mines, Trans. Roy. Soc. Can. {\bf 4}, 43 (1914).

\end{thebibliography}
\end{document}